\documentclass[twocolumn]{autart}    
\usepackage[T1]{fontenc}
\usepackage{graphicx}
\usepackage{amsmath,amssymb}
\usepackage{amssymb}
\usepackage{amsfonts}
\usepackage{hyperref}
\usepackage{xcolor}
\usepackage{bm}
\newtheorem{assumption}{Assumption}
\newtheorem{definition}{Definition}
\newtheorem{proposition}{Proposition}

\newtheorem{remark}{Remark}
\newtheorem{theorem}{Theorem}
\newcommand{\oomit}[1]{}

\begin{document}

\begin{frontmatter}

\title{Reachability Verification for Stochastic Discrete-time Dynamical Systems} 

\author[Paestum]{Bai Xue}\ead{xuebai@ios.ac.cn}    

\address[Paestum]{State Key Laboratory of Computer Science, Institute of Software, CAS}


\begin{abstract}                         
In this paper we study reachability verification problems of stochastic discrete-time dynamical systems over the infinite time horizon. The reachability verification of interest in this paper is to certify specified lower and upper bounds of the reachability probability, with which the system starting from a designated initial set will enter a desired target set eventually. Existing sufficient conditions for reachability verification over the infinite time horizon are established based on the Doob's non-supermartingle inequality, which are often restrictive. Recently, a set of equations was proposed in \cite{xue2021reach}, to which the solution is able to characterize the exact reachability probability. Inspired by this result, we in this paper propose sufficient conditions for reachability verification based on relaxing this equation. These sufficient conditions are shown to be weaker than the state-of-the-art ones, consequently being capable of providing more opportunities for verifying the reachability property successfully.
\end{abstract}

\end{frontmatter}

\section{Introduction}
\label{sec:intro}
In classical analysis, complex models, such as stochastic difference equations, are usually checked against simple specifications. Examples include the stability of an equilibrium, the invariance of a set, and properties such as controllability and observability \cite{khalil2002nonlinear}. There is a growing interest, however, in using formal methods to check the behavior of a complex model against rich temporal specifications that include notions of safety (i.e., something bad never happens) and its dual, reachability (i.e., something good eventually happens safely).  

A popular approach to temporal verification in deterministic systems is via barrier functions which provide Lyapunov-like guarantees regarding system behavior. The existence of a barrier function is enough to conclude the satisfiability of safety or reachability specifications \cite{prajna2007framework,prajna2007convex}. Later, significant efforts have been devoted to modifying and improving the deterministic form of barrier functions as well as expanding their applications \cite{ames2019control,xue2022reach}. However, many real-world applications are subject to stochastic disturbances and are modeled as stochastic systems. In the stochastic setting, safety verification  over the infinite time horizon via barrier certificates was introduced in \cite{prajna2007framework} along with the deterministic counterpart. Based on the Doob's supermartingale inequality \cite{oksendal2013stochastic}, \cite{prajna2007framework} constructed a non-negative barrier function and provided a sufficient condition for certifying specified upper bounds of probabilities, with which a system starting from a specified initial set will enter an unsafe region. Unfortunately, the formulated sufficient condition can be overly restrictive because it requires the expected value evolution of the barrier function to be monotonically non-increasing, i.e., the barrier function is restricted to be a non-negative supermartingale. In order to alleviate this issue, the $c$-martingale was proposed, which permits a bounded increase in the expected value of the certificate at each time step. However, the probabilistic guarantee it produces can only be established by bounded time horizons \cite{steinhardt2012finite,santoyo2021barrier}. Afterwards, inspired by the $k$-inductive principle \cite{sheeran2000checking}, new sufficient conditions for safety and reachability verification were proposed to further alleviate this issue in \cite{anand2022k}. Essentially, the construction of these sufficient conditions relies on the Doob's supermartingale inequality as well. Very recently, a sufficient condition for inner-approximate reachability analysis, which is constructed by relaxing a set of equations being able to characterize the exact reachability probability, was formulated in \cite{xue2021reach}. The inner-approximate reachability analysis is inner-approximating the set of all initial states, starting from each of which the system will enter a desired target set with a probability being larger than a specified threshold. The formulated sufficient condition can be used for certifying lower bounds of reachability probabilities via supplementing a requirement that the designated initial set is included in the computed inner-approximation. This method is orthogonal to the aforementioned methods based on the Doob's supermartingale inequality, and deserves further exploration in reachability verification.

The present work studies the reachability verification problem of stochastic discrete-time systems over the infinite time horizon based on relaxing the set of equations in \cite{xue2021reach}. The reachability verification problem of interest in this paper is to certify both lower and upper bounds of the reachability probability, with which the system starting from a specified initial set will enter a desired target set eventually. Firstly, we complement the results in  \cite{xue2021reach} and present sufficient conditions for certifying an upper bound of the reachability probability. Secondly, we extend these conditions and further propose sufficient conditions for reachability verification with the $k$-induction principle. The comparison between the proposed ones and the state-of-the-art ones demonstrates that our proposed conditions are more expressive and can provide more chances of verifying the reachability specification successfully.

\subsection*{Related Work}
\label{sec:rw}
Verification of dynamical systems against complex temporal specifications has gained increasing attention in the last few years \cite{lavaei2022automated}. Given desired temporal properties, formal verification is concerned to soundly check whether these properties are satisfied. In case that the underlying systems are subject to stochastic uncertainties, the objective turns into formally quantifying the probability of satisfying the property of interest \cite{baier2008principles}. However, verification for complex temporal specifications such as reachability for these systems with continuous state spaces is generally intractable to solve. Existing methods are mainly categorized into discretization-based and discretization-free ones \cite{lavaei2022automated}.

Discretization-based  methods typically involve the discretization of the system's domain into a finite number of discrete states, resulting in a finite stochastic transition system. This transition system serves as a finite-state abstraction of the continuous-state dynamics. Performing verification on  this abstraction is generally more tractable and yields bounded-error probabilistic guarantees with respect to the original system states. As a result, several types of stochastic abstractions, such as approximate markov chains 
 \cite{tkachev2011infinite,tkachev2014characterization} and interval-valued markov chains \cite{lahijanian2015formal,chatterjee2008model}, have been put forth in the literature. However, these abstraction techniques face the issue of discrete state exploration. This critical challenge motivates the development of discretization-free approaches.

Besides the methods based on satisfiability modulo theory \cite{franzle2008stochastic}, one well-known discretization-free method is the barrier certificates method \cite{prajna2007framework}. Barrier certificates are Lyapunov-like functions defined over the state space of the system and satisfying a set of inequalities on both the function itself and the one-step transition (or the infinitesimal generator along the flow) of the system. An appropriate level set of a barrier certificate can separate an unsafe region from all system trajectories starting from a given set of initial states with some probability lower bound. Consequently, the existence of such a function provides a formal probabilistic certificate for system safety and can certify upper bounds of the probability of reaching unsafe sets. Recently, inspired by the results in \cite{prajna2007convex}, a modified barrier certificate, which is able to certify lower bounds of the probability of reaching unsafe or target sets, was proposed in  \cite{anand2022k}. However, in order to provide infinite time horizon guarantees, all of these results require an assumption that the barrier function should be a non-negative supermartingale, which is rather restrictive in practice. In order to alleviate this issue, new results, termed $k$-inductive barrier certificates, were further developed for certifying both lower and upper probability bounds based on the $k$-inductive principle in \cite{anand2022k}. The $k$-inductive barrier certificate relaxes the classical non-negative supermartingale based barrier certificate by permitting an increase in the expected value of the certificate at some time steps. It is a barrier certificate for $k$-compositions of the system. Nevertheless, the construction of $k$-inductive barrier certificates also relies on the Doob’s supermartingale inequality. 

Like the work \cite{anand2022k}, the present work also studies reachability verification of stochastic discrete-time dynamical systems. However, the results developed in this paper do not rely on the Doob’s supermartingale inequality. They are established based on relaxing the set of equations in \cite{xue2021reach}. Compared to the ones in \cite{anand2022k}, the resulting certificates are not required to be non-negative and are shown to be more powerful in reachability verification.   

This paper is structured as follows. In Section \ref{sec:pre} we introduce the reachability verification problem of interest and recall existing sufficient conditions. We first present a sufficient condition for certifying upper bounds of the reachability probability via relaxing the set of equations in \cite{xue2021reach} directly in Section \ref{sec:comp}, and then in Section \ref{k_R_V} present new sufficient conditions with the $k$-inductive principle. Finally, we conclude this paper in Section \ref{sec:con}.
\section{Preliminaries}
\label{sec:pre}
We start our exposition by a formal introduction of discrete-time systems subject to stochastic disturbances and reachability verification problems of interest. Before posing the problem studied, let us introduce some basic notions used throughout this paper: $\mathbb{N}$ denotes the set of nonnegative integers; for a set $\Delta$, $\Delta^c$ and $\partial \Delta$ denote the complement and the boundary of the set $\Delta$, respectively; $\mathbb{R}_{\geq 0}$ is the set of non-negative real numbers;  $1_A(\bm{x})$ denotes the indicator function in the set $A$,
where, if $\bm{x}\in A$, then $1_A(\bm{x}) = 1$ and if $\bm{x}\notin A$, $1_A(\bm{x}) = 0$. 

\subsection{Problem Statement}
In this paper we consider stochastic discrete-time systems that can be modeled by stochastic difference equations of the following form:
\begin{equation}
\label{system}
\begin{split}
&\bm{x}(l+1)=\bm{f}(\bm{x}(l),\bm{\theta}(l)), \forall l\in \mathbb{N},\\
&\bm{x}(0)=\bm{x}_0\in \mathcal{X},
\end{split}
\end{equation}
where $\bm{x}(\cdot): \mathbb{N} \rightarrow \mathbb{R}^n$ are states, and $\bm{\theta}(\cdot): \mathbb{N}\rightarrow \Theta$ with $\Theta \subseteq \mathbb{R}^m$ are stochastic disturbances. In addition, suppose that the random vectors, $\bm{\theta}(0), \bm{\theta}(1),\ldots$, are independent and identically distributed (i.i.d), and take values in  $\Theta$ with the following probability distribution,
\[{\rm Prob}(\bm{\theta}(l)\in B)=\mathbb{P}(B), \forall l\in \mathbb{N}, \forall B\subseteq \Theta.\] 
Correspondingly, $\mathbb{E}[\cdot]$ is the expectation induced by the distribution $\mathbb{P}$.

Let $\Theta\times \Theta=\Theta^2$. Then, the 2-composition of system \eqref{system} is given by 
\[\bm{x}(l+2)=\bm{f}(\bm{f}(\bm{x}(l),\bm{\theta}(l)),\bm{\theta}(l+1)):=\bm{f}^2(\bm{x}(l),\bm{\theta}(l:l+1)),\]
where $\bm{f}^2: \mathbb{R}^n \times \Theta^2\rightarrow \mathbb{R}^n$ and $\bm{\theta}({l}:{l+1})\in \Theta^2$. Since the sequence of random vectors $\{\bm{\theta}(l),l\in \mathbb{N}\}$ is assumed i.i.d, the probability measure on $\Theta^2$ will simply be the product measure, i.e., $\mathbb{P}\times \mathbb{P}:=\mathbb{P}^2$. Similarly, the $(l+1)$-composition is denoted by $\bm{x}(i+1)=\bm{f}^{l+1}(\bm{x}(i),\bm{\theta}(li:li+l))$, where $\bm{f}^{l+1}: \mathbb{R}^n\times \Theta^{l+1}\rightarrow \mathbb{R}^n$ and $\bm{\theta}(li:li+l)\in \Theta^{l+1}$ with the probability measure $\mathbb{P}^{l+1}$.

Before defining the trajectory of system \eqref{system}, we define a disturbance signal.
\begin{definition}
A disturbance signal $\pi$ is an ordered sequence $\{\bm{\theta}(i),i\in \mathbb{N}\}$, where $\bm{\theta}(\cdot): \mathbb{N}\rightarrow \Theta$.
\end{definition}

Given system \eqref{system}, a signal $\pi=\{\bm{\theta}(i),i\in \mathbb{N}\}$ is a stochastic process defined on the canonical sample space $\Omega=\Theta^{\infty}$, endowed with its product topology $\mathcal{B}(\Omega)$, with the probability measure $\mathbb{P}^{\infty}$.   The expectation associated with the probability measure $\mathbb{P}^{\infty}$ is denoted by $\mathbb{E}^{\infty}[\cdot]$.

A disturbance signal $\pi$ together with an initial state $\bm{x}_0\in \mathbb{R}^n$ induces a unique discrete-time trajectory as follows.
\begin{definition}
Given a disturbance signal $\pi \in \Omega$ and an initial state $\bm{x}_0\in \mathbb{R}^n$, a trajectory of system \eqref{system} is denoted as  $\bm{\phi}_{\pi}^{\bm{x}_0}(\cdot):\mathbb{N}\rightarrow \mathbb{R}^n$ with $\bm{\phi}_{\pi}^{\bm{x}_0}(0)=\bm{x}_0$, i.e.,
\begin{equation*}
\bm{\phi}_{\pi}^{\bm{x}_0}(l+1)=\bm{f}(\bm{\phi}_{\pi}^{\bm{x}_0}(l),\bm{\theta}(l)), \forall l\in \mathbb{N}.
\end{equation*}
\end{definition}

Given a state constraint set $\mathcal{X}\subseteq \mathbb{R}^n$ satisfying Assumption \ref{asump}, an initial set $\mathcal{X}_0$ and a target set $\mathcal{X}_r$, where $\mathcal{X}_0,\mathcal{X}_r\subseteq \mathcal{X}$, the reachability verification is to certify lower and upper bounds on the probability of reaching the target set $\mathcal{X}_r$ eventually for system \eqref{system}, starting from the initial set $\mathcal{X}_0$.

\begin{assumption}
\label{asump}
   $\bm{f}(\bm{x},\bm{\theta}):\mathcal{X}\times \Theta \rightarrow \mathcal{X}$, i.e., for any solution process $\bm{\phi}^{\bm{x}_0}_{\pi}(\cdot): \mathbb{N}\rightarrow \mathbb{R}$ starting from $\bm{x}_0\in \mathcal{X}$, we have  $\bm{\phi}^{\bm{x}_0}_{\pi}(k)\in \mathcal{X}$ for all $k\in \mathbb{N}$. If $\mathcal{X}=\mathbb{R}^n$, this assumption is meaningless.
\end{assumption}

\begin{definition}
\label{ravoid}
Given thresholds $\epsilon_1 \in [0,1]$ and $\epsilon_2\in [0,1]$, the reachability verification problem is to certify whether $\epsilon_1$ and $\epsilon_2$ are respectively the lower and upper bounds of the probability, with which system \eqref{system} starting from each state in the initial set $\mathcal{X}_0$ will enter the target set $\mathcal{X}_r$ eventually, i.e., to certify 
\[\epsilon_1\leq \mathbb{P}^{\infty}\Big(\exists k\in \mathbb{N}. \bm{\phi}_{\pi}^{\bm{x}_0}(k)\in \mathcal{X}_r\mid \bm{x}_0\in \mathcal{X}_0\Big)\leq \epsilon_2.\]
\end{definition}

\begin{remark}
 If Assumption \ref{asump} is discarded, the reachability verification problem in Definition \ref{ravoid} turns into certifying whether $\epsilon_1$ and $\epsilon_2$ are respectively the lower and upper bounds of the probability, with which system \eqref{system} starting from each state in the initial set $\mathcal{X}_0$ will enter the target set $\mathcal{X}_r$ eventually while staying inside the set $\mathcal{X}$ before the first target hitting time, i.e., certifying 
 \[\epsilon_1\leq \mathbb{P}^{\infty}\left(
\begin{aligned}
&\exists k\in \mathbb{N}. \bm{\phi}_{\pi}^{\bm{x}_0}(k)\in \mathcal{X}_r
\bigwedge \\
&\forall l\in [0,k]\cap \mathbb{N}. \bm{\phi}_{\pi}^{\bm{x}_0}(l)\in \mathcal{X}
\end{aligned}
\mid \bm{x}_0\in \mathcal{X}_0
\right)\leq \epsilon_2.\]

In order to make fair comparisons with existing ones (i.e., Proposition \ref{the:barrier}-\ref{k-inductive_reachability} in Subsection \ref{doob}), in this paper we illustrate our results under Assumption \ref{asump}. Some results without Assumption \ref{asump} are also presented in this paper.
\end{remark}

\subsection{Reachability Verification Based on the Doob's Supermartingale Inequality}
\label{doob}
In this section we recall sufficient conditions for reachability verification  based on the Doob's supermartingale inequality.

A supermartingale is a sequence of random variables for which the conditional expectation of the next value in the sequence is smaller than or equal to the present value irrespective of the prior values. For nonnegative supermartingales, there exists the following well-known result -- the Doob's supermartingale inequality.

\begin{theorem}
    Let $(\Omega,\mathcal{F},\mathbb{P}_1)$ be the probability space and $(B_i)_{i\in \mathbb{N}}$ be an non-negative supermartingale, then for $b>0$, we have that 
    \[\mathbb{P}_1\big(\sup_{i\in \mathbb{N}}B_i\geq b\mid B_0 \big)\leq \frac{B_0}{b}.\]
\end{theorem}

Sufficient conditions were proposed for reachability verification in the sense of Definition \ref{ravoid}. A sufficient condition for certifying upper bounds of the reachability probability  is formulated in Proposition \ref{the:barrier}. 

\begin{proposition} [Theorem 5, \cite{anand2022k}]
\label{the:barrier}
 Under Assumption \ref{asump},  let $v(\bm{x}): \mathcal{X}\rightarrow \mathbb{R}_{\geq 0}$ be a barrier certificate satisfying 
    \begin{equation}
    \label{sbs}
        \begin{cases}
            &v(\bm{x})\leq \epsilon_2, \forall \bm{x}\in \mathcal{X}_0,\\
            &v(\bm{x})\geq 1, \forall \bm{x} \in \mathcal{X}_r,\\
            &\mathbb{E}^{\infty}[v(\bm{\phi}^{\bm{x}}_{\pi}(1))]-v(\bm{x}) \leq 0, \forall \bm{x} \in \mathcal{X},
        \end{cases}
    \end{equation}
then $\mathbb{P}^{\infty}\Big(\exists k\in \mathbb{N}. \bm{\phi}_{\pi}^{\bm{x}}(k)\in \mathcal{X}_r\mid \bm{x}\in \mathcal{X}_0\Big)\leq \epsilon_2$.  
\end{proposition}

If the set $\mathcal{X}_r$ is an unsafe set, then a function $v(\bm{x})$ satisfying constraint \eqref{sbs} in Proposition \ref{the:barrier} is the well-known barrier certificate in existing literature for safety verification. Under Assumption \ref{asump}, the Doob's supermartingale inequality also facilitates the construction of sufficient conditions for certifying lower bounds of the reachability probability.

\begin{proposition}[Theorem 16, \cite{anand2022k}]
\label{reachability_existing}
   Under Assumption \ref{asump}, let $v(\bm{x}): \mathcal{X}\rightarrow \mathbb{R}_{\geq 0}$ be a barrier certificate satisfying 
    \begin{equation}
    \label{super_reach}
        \begin{cases}
            &v(\bm{x})\leq 1-\epsilon_1, \forall \bm{x}\in \mathcal{X}_0,\\
            &v(\bm{x})\geq 1, \forall \bm{x} \in \partial \mathcal{X}\setminus \partial \mathcal{X}_r,\\
            &\mathbb{E}^{\infty}[v(\bm{\phi}_{\pi}^{\bm{x}}(1))]-v(\bm{x}) \leq -\delta, \forall \bm{x} \in \overline{\mathcal{X}\setminus \mathcal{X}_r},
        \end{cases}
    \end{equation}
   where $\delta > 0$ is a user-defined value, then $\mathbb{P}^{\infty}\big(\exists k\in \mathbb{N}. \bm{\phi}^{\bm{x}}_{\pi}(k) \in \mathcal{X}_r\mid \bm{x}\in \mathcal{X}_0\big)\geq \epsilon_1$. 
\end{proposition}

\begin{remark}
Another condition, which is similar to the one in Proposition \ref{reachability_existing} but provides stronger almost-sure guarantees, was proposed in \cite{chakarov2013probabilistic}. 
\begin{proposition}
    Under Assumption  \ref{asump}, if there exist a function  $v(\bm{x}): \mathcal{X}\rightarrow \mathbb{R}_{\geq 0}$ and constant $c>0$ such that 
    \begin{equation}
        \begin{cases}
            &v(\bm{x})\geq c, \forall \bm{x}\in \mathcal{X}\setminus \mathcal{X}_r,\\
            &v(\bm{x})<c, \forall \bm{x}\in \mathcal{X}_r,\\
            &\mathbb{E}^{\infty}[v(\bm{\phi}^{\bm{x}}_{\pi}(1))]-v(\bm{x})\leq -1, \forall \bm{x}\in \mathcal{X}\setminus \mathcal{X}_r,
        \end{cases}
    \end{equation}
    then $\mathbb{P}^{\infty}\big(\exists k\in \mathbb{N}. \bm{\phi}^{\bm{x}}_{\pi}(k) \in \mathcal{X}_r\mid \bm{x}\in \mathcal{X}_0\big)=1$. 
\end{proposition}
\end{remark}
 
Meanwhile, in \cite{anand2022k} $k$-inductive barrier certificates, which relax the non-negative supermartingale based barrier certificates in Proposition \ref{the:barrier} and \ref{reachability_existing} by permitting an
increase in the expected value of the certificate at some times, were also proposed for reachability verification.

\begin{proposition}[Theorem 11, \cite{anand2022k}]
    \label{k-inductive_safe}
    Under Assumption \ref{asump}, let $v(\bm{x}):\mathcal{X}\rightarrow \mathbb{R}_{\geq 0}$ be a k-inductive barrier certificate satisfying
    \begin{equation}
    \label{k-inductive_safe1}
        \begin{cases}
            &v(\bm{x})\leq \epsilon'_2,\forall \bm{x}\in \mathcal{X}_0,\\
            &v(\bm{x})\geq 1, \forall \bm{x}\in \mathcal{X}_r,\\
            &\mathbb{E}^{\infty}[v(\bm{\phi}_{\pi}^{\bm{x}}(1))]-v(\bm{x})\leq c, \forall \bm{x}\in \mathcal{X},\\
            &\mathbb{E}^{\infty}[v(\bm{\phi}_{\pi}^{\bm{x}}(k))]-v(\bm{x})\leq 0, \forall \bm{x}\in \mathcal{X},
        \end{cases}
    \end{equation}
    where $c\geq 0$ is a user-defined value and $\epsilon'_2\in [0,1]$, then $\mathbb{P}^{\infty}\big(\exists k\in \mathbb{N}. \bm{\phi}^{\bm{x}}_{\pi}(k) \in \mathcal{X}_r\mid \bm{x}\in \mathcal{X}_0\big)\leq k\epsilon'_2+\frac{k(k-1)c}{2}$. 
\end{proposition}

By setting $\epsilon'_2:=\frac{2\epsilon_2-k(k-1)c}{2k}$ in Proposition \ref{k-inductive_safe}, we have $\mathbb{P}^{\infty}\big(\exists k\in \mathbb{N}. \bm{\phi}^{\bm{x}}_{\pi}(k) \in \mathcal{X}_r\mid \bm{x}\in \mathcal{X}_0\big)\leq \epsilon_2$ if constraint \eqref{k-inductive_safe1} holds.

\begin{proposition}[Theorem 22, \cite{anand2022k}]
    \label{k-inductive_reachability}
    Under Assumption \ref{asump}, let $v(\bm{x}):\mathcal{X}\rightarrow \mathbb{R}_{\geq 0}$ be a k-inductive barrier certificate satisfying
    \begin{equation}
    \label{k-inductive_reachability1}
        \begin{cases}
            &v(\bm{x})\leq \epsilon'_1,\forall \bm{x}\in \mathcal{X}_0,\\
            &v(\bm{x})\geq 1, \forall \bm{x}\in \partial \mathcal{X}\setminus \partial \mathcal{X}_r,\\
            &\mathbb{E}^{\infty}[v(\bm{\phi}_{\pi}^{\bm{x}}(1))]-v(\bm{x})\leq c, \forall \bm{x}\in \overline{\mathcal{X}\setminus \mathcal{X}_r},\\
            &\mathbb{E}^{\infty}[v(\bm{\phi}_{\pi}^{\bm{x}}(k))]-v(\bm{x})\leq -\delta, \forall \bm{x}\in \overline{\mathcal{X}\setminus \mathcal{X}_r},
        \end{cases}
    \end{equation}
    where $c\geq 0$ is a user-defined value and $\epsilon'_1\in [0,1]$, then $\mathbb{P}^{\infty}\big(\exists k\in \mathbb{N}. \bm{\phi}^{\bm{x}}_{\pi}(k) \in \mathcal{X}_r\mid \bm{x}\in \mathcal{X}_0\big)\geq 1-k\epsilon'_1-\frac{k(k-1)c}{2}$. 
\end{proposition}

By setting $\epsilon'_1:=\frac{2-k(k-1)c-2\epsilon_1}{2k}$ in Proposition \ref{k-inductive_reachability}, we have that
$\mathbb{P}^{\infty}\big(\exists k\in \mathbb{N}. \bm{\phi}^{\bm{x}}_{\pi}(k) \in \mathcal{X}_r\mid \bm{x}\in \mathcal{X}_0\big)\geq \epsilon_1$ if constraint \eqref{k-inductive_reachability1} holds.

\subsection{Reachability Verification Based on Equation Relaxations in \cite{xue2021reach}}
In this subsection we introduce another method of constructing sufficient conditions for reachability verification. This method does not need Assumption \ref{asump} and the Doob's supermartingale inequality. In contrast, this method is based on the relaxation of a set of equations, the solution to which is able to characterize the exact reachability probability.   Recently, it was extended to continuous-time systems modelled by stochastic differential equations \cite{xue2022reach1}. 
\begin{theorem}[Theorem 1, \cite{xue2021reach}]
\label{theorem_reach}
   If there exist bounded functions $v(\bm{x}):\widehat{\mathcal{X}}\rightarrow \mathbb{R}$ and $w(\bm{x}): \widehat{\mathcal{X}}\rightarrow \mathbb{R}$ such that for $\bm{x}\in \widehat{\mathcal{X}}$, 
   \begin{equation}
   \label{reach_equa}
   \begin{cases}
       &v(\bm{x})=\mathbb{E}^{\infty}[v(\widehat{\bm{\phi}}^{\bm{x}}_{\pi}(1))],\\
       &v(\bm{x})=1_{\mathcal{X}_r}(\bm{x})+\mathbb{E}^{\infty}[w(\widehat{\bm{\phi}}^{\bm{x}}_{\pi}(1))]-w(\bm{x}),
   \end{cases}
   \end{equation}
   then $\mathbb{P}^{\infty}\big(\exists k\in \mathbb{N}. \bm{\phi}^{\bm{x}_0}_{\pi}(k) \in \mathcal{X}_r\bigwedge \forall l\in [0,k]. \bm{\phi}^{\bm{x}_0}_{\pi}(l)\in \mathcal{X} \mid  \bm{x}_0\in \mathcal{X} \big)=\mathbb{P}^{\infty}\big(\exists k\in \mathbb{N}. \widehat{\bm{\phi}}^{\bm{x}_0}_{\pi}(k) \in \mathcal{X}_r \mid  \bm{x}_0\in \mathcal{X} \big)=v(\bm{x})=\lim_{i\rightarrow \infty}\frac{1_{\mathcal{X}_r}(\bm{x}_0)+\ldots+\mathbb{E}^{\infty}[1_{\mathcal{X}_r}(\widehat{\bm{\phi}}^{\bm{x}_0}_{\pi}(i-1))]}{i}$, where 
$\widehat{\bm{\phi}}^{\bm{x}_0}_{\pi}(\cdot): \mathbb{N}\rightarrow \mathbb{R}^n$ is the trajectory to the system 
\begin{equation}
\label{new_system}
\begin{cases}
&\bm{x}(j+1)=1_{\mathcal{X}\setminus \mathcal{X}_r}(\bm{x}(j))\cdot \bm{f}(\bm{x}(j),\bm{\theta}(j))\\
&+1_{\mathcal{X}_r}(\bm{x}(j))\cdot \bm{x}(j)+1_{\widehat{\mathcal{X}}\setminus \mathcal{X}}(\bm{x}(j))\cdot \bm{x}(j), \forall j\in \mathbb{N},\\
&\bm{x}(0)=\bm{x}_0,
\end{cases}
\end{equation}
and  $\widehat{\mathcal{X}}$ is a set satisfying $\widehat{\mathcal{X}}\supseteq \{\bm{x}\in \mathbb{R}^n\mid \bm{x}=\bm{f}(\bm{x}_0,\bm{\theta}), \bm{x}_0\in \mathcal{X}, \bm{\theta}\in \Theta\}\cup \mathcal{X}$.
\end{theorem}

Via relaxing the set of equations \eqref{reach_equa}, a set of inequalities is  obtained and the $p$-super level set (i.e., $\{\bm{x}\in \mathcal{X}\mid v(\bm{x})\geq p\}$) of its solution 
is a set of initial states (i.e., an inner-approximation of the $p$-reach-avoid set), from which system \eqref{system} starting will enter the target set $\mathcal{X}_r$ eventually while staying inside the set $\mathcal{X}$ preceding the target hit with a probability being larger than or equal to $p$. Please refer to Corollary 2 in \cite{xue2021reach}. By incorporating the constraint that the initial set $\mathcal{X}_0$ is a subset of the computed $p$-reach-avoid set, a sufficient condition can be obtained straightforwardly for certifying lower bounds of the reachability probability.

\begin{proposition}
\label{reachability}
   If there exist bounded functions $v(\bm{x}):\widehat{\mathcal{X}}\rightarrow \mathbb{R}$ and $w(\bm{x}): \widehat{\mathcal{X}}\rightarrow \mathbb{R}$ such that 
   \begin{equation*}
   \label{rb}
   \begin{cases}
       &v(\bm{x}) \geq \epsilon_1, \forall \bm{x}\in \mathcal{X}_0,\\
       &v(\bm{x})\leq \mathbb{E}^{\infty}[v(\widehat{\bm{\phi}}^{\bm{x}}_{\pi}(1))], \forall \bm{x}\in \widehat{\mathcal{X}},\\
       &v(\bm{x})\leq 1_{\mathcal{X}_r}(\bm{x})+\mathbb{E}^{\infty}[w(\widehat{\bm{\phi}}^{\bm{x}}_{\pi}(1))]-w(\bm{x}), \forall \bm{x}\in \widehat{\mathcal{X}},
   \end{cases}
   \end{equation*}
   which is equivalent to 
      \begin{equation}
   \label{rb1}
   \begin{cases}
       &v(\bm{x})\geq \epsilon_1,\forall \bm{x}\in \mathcal{X}_0,\\ 
       &v(\bm{x})\leq \mathbb{E}^{\infty}[v(\bm{\phi}^{\bm{x}}_{\pi}(1))], \forall \bm{x}\in \mathcal{X}\setminus \mathcal{X}_r,\\
       &v(\bm{x})\leq \mathbb{E}^{\infty}[w(\bm{\phi}^{\bm{x}}_{\pi}(1))]-w(\bm{x}), \forall \bm{x}\in \mathcal{X}\setminus \mathcal{X}_r,\\
       &v(\bm{x})\leq 1, \forall \bm{x}\in \mathcal{X}_r,\\
       &v(\bm{x}) \leq 0, \forall \bm{x} \in \widehat{\mathcal{X}} \setminus \mathcal{X},
   \end{cases}
   \end{equation}
   then $\mathbb{P}^{\infty}\big(\exists k\in \mathbb{N}. \bm{\phi}^{\bm{x}}_{\pi}(k) \in \mathcal{X}_r\bigwedge \forall l\in [0,k]. \bm{\phi}^{\bm{x}}_{\pi}(l)\in \mathcal{X} \mid  \bm{x}\in \mathcal{X}_0 \big)=\mathbb{P}^{\infty}\big(\exists k\in \mathbb{N}. \widehat{\bm{\phi}}^{\bm{x}}_{\pi}(k) \in \mathcal{X}_r \mid  \bm{x}\in \mathcal{X}_0 \big)\geq \epsilon_1$.  
\end{proposition}

Theorem \ref{theorem_reach} and Proposition \ref{reachability} do not need Assumption \ref{asump}. Instead, in order to maintain invariance, an auxiliary set $\widehat{\mathcal{X}}$, which is an invariant of the switched system \eqref{new_system}, is required. In the following we will further construct new sufficient conditions for reachability verification based on relaxing the set of equations \eqref{reach_equa}.  Under Assumption \ref{asump}, the set $\widehat{\mathcal{X}}$ is not needed anymore.

\section{Reachability Verification}
\label{sec:comp}
Under Assumption \ref{asump}, let’s first compare constraints \eqref{super_reach} and \eqref{rb1}, motivating the use of relaxations of the set of equations \eqref{reach_equa} for reachability verification. In this context, due to the fact that $\widehat{\mathcal{X}}$ is not needed any more, constraint \eqref{rb1} turns into 
 \begin{equation}
   \label{rb2}
   \begin{cases}
       &v(\bm{x})\geq \epsilon_1,\forall \bm{x}\in \mathcal{X}_0,\\ 
       &v(\bm{x})\leq \mathbb{E}^{\infty}[v(\bm{\phi}^{\bm{x}}_{\pi}(1))], \forall \bm{x}\in \mathcal{X}\setminus \mathcal{X}_r,\\
       &v(\bm{x})\leq \mathbb{E}^{\infty}[w(\bm{\phi}^{\bm{x}}_{\pi}(1))]-w(\bm{x}), \forall \bm{x}\in \mathcal{X}\setminus \mathcal{X}_r,\\
       &v(\bm{x})\leq 1, \forall \bm{x}\in \mathcal{X}_r.
   \end{cases}
   \end{equation}
  Via setting $u(\bm{x}):=1-v(\bm{x})$, constraint \eqref{super_reach} with $v(\bm{x})\geq 0$ over $\mathcal{X}$ can be reformulated as the following equivalent form 
    \begin{equation}
    \label{super_reach1}
        \begin{cases}
            &u(\bm{x})\geq \epsilon_1, \forall \bm{x}\in \mathcal{X}_0,\\
            &u(\bm{x})\leq 0, \forall \bm{x} \in \partial \mathcal{X}\setminus \partial \mathcal{X}_r,\\
            &\mathbb{E}^{\infty}[u(\bm{\phi}_{\pi}^{\bm{x}}(1))]-u(\bm{x}) \geq \delta, \forall \bm{x} \in \overline{\mathcal{X}\setminus \mathcal{X}_r},\\
            &u(\bm{x})\leq 1, \forall \bm{x}\in \mathcal{X}.
        \end{cases}
    \end{equation}
We will show that
if a bounded function $v(\bm{x})$ satisfies \eqref{super_reach1}, it will satisfy \eqref{rb2}.  We just need to show that if there exists a bounded function $v(\bm{x})$ satisfying
\[\mathbb{E}^{\infty}[v(\bm{\phi}_{\pi}^{\bm{x}_0}(1))]-v(\bm{x})\geq \delta, \forall \bm{x}\in \overline{\mathcal{X}\setminus \mathcal{X}_r},\]
there exists a bounded function $w(\bm{x}):\mathcal{X}\rightarrow \mathbb{R}$ such that 
\begin{equation*}
    \begin{cases}
         &v(\bm{x})\leq \mathbb{E}^{\infty}[v(\bm{\phi}^{\bm{x}}_{\pi}(1))], \forall \bm{x}\in \mathcal{X}\setminus \mathcal{X}_r,\\
       &v(\bm{x})\leq \mathbb{E}^{\infty}[w(\bm{\phi}^{\bm{x}}_{\pi}(1))]-w(\bm{x}), \forall \bm{x}\in \mathcal{X}\setminus \mathcal{X}_r.
    \end{cases}
\end{equation*}
Obviously, if $v(\bm{x})$ satisfies 
\[
\mathbb{E}^{\infty}[v(\bm{\phi}_{\pi}^{\bm{x}}(1))]-v(\bm{x}))\geq \delta, \forall \bm{x}\in \overline{\mathcal{X}\setminus \mathcal{X}_r},
\]
it satisfies \[v(\bm{x}) \leq  \mathbb{E}^{\infty}[v(\bm{\phi}_{\pi}^{\bm{x}}(1)], \forall \bm{x} \in \mathcal{X}\setminus \mathcal{X}_r.\] Further, via taking $w(\bm{x}) =Mv(\bm{x})$, where $M>0$ satisfies $M \delta \geq \sup_{\bm{x}\in \mathcal{X}} v(\bm{x})$, we have \[v(\bm{x})\leq \mathbb{E}^{\infty}[w(\bm{\phi}_{\pi}^{\bm{x}}(1))]-w(\bm{x}), \forall \bm{x}\in \mathcal{X}\setminus \mathcal{X}_r.\] Thus, the conclusion holds and constraint \eqref{rb2} is weaker than constraint \eqref{super_reach1}, providing more possibilities of verifying that $\epsilon_1$ is a lower bound of the reachability probability successfully.

Besides the gain of a sufficient condition for certifying lower bounds of the reachability probability via relaxing the set of equations \eqref{reach_equa}, a sufficient condition can also be obtained for certifying an upper bound of the reachability probability. This sufficient condition is formulated in Proposition \ref{SB}.
\begin{proposition}
\label{SB}
Under Assumption \ref{asump}, if there exist bounded functions $v(\bm{x}):\mathcal{X}\rightarrow \mathbb{R}$ and $w(\bm{x}): \mathcal{X}\rightarrow \mathbb{R}$ such that 
   \begin{equation}
   \label{sb1}
   \begin{cases}
      &v(\bm{x})\leq \epsilon_2,\forall \bm{x}\in \mathcal{X}_0,\\ 
       &v(\bm{x})\geq \mathbb{E}^{\infty}[v(\bm{\phi}^{\bm{x}}_{\pi}(1))], \forall \bm{x}\in \mathcal{X}\setminus \mathcal{X}_r,\\
       &v(\bm{x})\geq \mathbb{E}^{\infty}[w(\bm{\phi}^{\bm{x}}_{\pi}(1))]-w(\bm{x}), \forall \bm{x}\in \mathcal{X}\setminus \mathcal{X}_r,\\
       &v(\bm{x})\geq 1, \forall \bm{x}\in \mathcal{X}_r,
   \end{cases}
   \end{equation}
   then $\mathbb{P}^{\infty}\big(\exists k\in \mathbb{N}. \bm{\phi}^{\bm{x}}_{\pi}(k) \in \mathcal{X}_r \mid  \bm{x}\in \mathcal{X}_0 \big)\leq \epsilon_2$.
\end{proposition}
\pf{
    From constraint \eqref{sb1}, we have
      \begin{equation*}
   \label{sb}
   \begin{cases}
      &v(\bm{x})\leq \epsilon_2,\forall \bm{x}\in \mathcal{X}_0,\\ 
       &v(\bm{x})\geq \mathbb{E}^{\infty}[v(\widetilde{\bm{\phi}}^{\bm{x}}_{\pi}(1))], \forall \bm{x}\in \mathcal{X},\\
       &v(\bm{x})\geq 1_{\mathcal{X}_r}(\bm{x})+\mathbb{E}^{\infty}[w(\widetilde{\bm{\phi}}^{\bm{x}}_{\pi}(1))]-w(\bm{x}), \forall \bm{x}\in \mathcal{X},
   \end{cases}
   \end{equation*}
where $\widetilde{\bm{\phi}}^{\bm{x}}_{\pi}(\cdot): \mathbb{N}\rightarrow \mathcal{X}$ is the trajectory to the system 
\begin{equation}
    \label{new_system1}
\begin{cases}
&\bm{x}(j+1)=1_{\mathcal{X}\setminus \mathcal{X}_r}(\bm{x}(j))\cdot \bm{f}(\bm{x}(j),\bm{\theta}(j))\\
&~~~~~~~~~~~~~~~~~~~~~~+1_{\mathcal{X}_r}(\bm{x}(j))\cdot \bm{x}(j), \forall j\in \mathbb{N},\\
&\bm{x}(0)=\bm{x}.
\end{cases}
\end{equation}

   Thus, we have that for $m\in \mathbb{N}$,
\begin{equation*}
    \begin{split}
    &v(\bm{x})\geq 1_{\mathcal{X}_r}(\bm{x})+\mathbb{E}^{\infty}[w(\widetilde{\bm{\phi}}^{\bm{x}}_{\pi}(1))]-w(\bm{x}),\\
    &v(\bm{x})\geq \mathbb{E}^{\infty}[v(\widetilde{\bm{\phi}}^{\bm{x}}_{\pi}(1))]\geq \mathbb{E}^{\infty}[1_{\mathcal{X}_r}(\widetilde{\bm{\phi}}^{\bm{x}}_{\pi}(1))]\\
    &~~~~~~~~~~~~~~~~~~~~~~~~~+\mathbb{E}^{\infty}[w(\widetilde{\bm{\phi}}^{\bm{x}}_{\pi}(2))]
    -\mathbb{E}^{\infty}[w(\widetilde{\bm{\phi}}^{\bm{x}}_{\pi}(1))],\\
    &\ldots,\\
    &v(\bm{x})\geq \mathbb{E}^{\infty}[v(\widetilde{\bm{\phi}}^{\bm{x}}_{\pi}(m))]\geq \mathbb{E}^{\infty}[1_{\mathcal{X}_r}(\widetilde{\bm{\phi}}^{\bm{x}}_{\pi}(m))]\\
    &~~~~~~~~~~~~~~~~~+\mathbb{E}^{\infty}[w(\widetilde{\bm{\phi}}^{\bm{x}}_{\pi}(m+1))]-\mathbb{E}^{\infty}[w(\widetilde{\bm{\phi}}^{\bm{x}}_{\pi}(m))]
    \end{split}
\end{equation*}
and thus 
\[
\begin{split}
v(\bm{x})&\geq \frac{\mathbb{E}^{\infty}[w(\widetilde{\bm{\phi}}^{\bm{x}}_{\pi}(m+1))]-w(\bm{x})}{m+1}\\
&+\frac{1_{\mathcal{X}_r}(\bm{x})+\ldots+\mathbb{E}^{\infty}[1_{\mathcal{X}_r}(\widetilde{\bm{\phi}}^{\bm{x}}_{\pi}(m))]}{m+1}, \forall m\in \mathbb{N},
\end{split}
\]
which implies, from Lemma 2 in \cite{xue2021reach},  
\[
\begin{split}
v(\bm{x})&\geq \lim_{m\rightarrow \infty} \frac{1_{\mathcal{X}_r}(\bm{x})+\ldots+\mathbb{E}^{\infty}[1_{\mathcal{X}_r}(\widetilde{\bm{\phi}}^{\bm{x}}_{\pi}(m))]}{m+1}\\
&=\mathbb{P}^{\infty}\big(\exists k\in \mathbb{N}. \widetilde{\bm{\phi}}^{\bm{x}}_{\pi}(k) \in \mathcal{X}_r\mid \bm{x}\in \mathcal{X}\big).
\end{split}
\]

    Also, since $\mathbb{P}^{\infty}\big(\exists k\in \mathbb{N}. \widetilde{\bm{\phi}}^{\bm{x}}_{\pi}(k) \in \mathcal{X}_r\mid \bm{x}\in \mathcal{X}_0\big)=\mathbb{P}^{\infty}\big(\exists k\in \mathbb{N}. \bm{\phi}^{\bm{x}}_{\pi}(k) \in \mathcal{X}_r\mid  \bm{x}\in \mathcal{X}_0 \big)$, and $v(\bm{x})\leq \epsilon_2, \forall \bm{x}\in \mathcal{X}_0$, we have the conclusion. \qed
}

From the proof of Proposition \ref{SB}, we can observe that for any state $\bm{x}$ in $\{\bm{x}\in \mathcal{X}\mid v(\bm{x})\leq \epsilon_2\}$, system \eqref{system} starting from it will enter the target set $\mathcal{X}_r$ eventually with a probability being smaller than or equal to $\epsilon_2$.

We compare constraints \eqref{sb1} and \eqref{sbs} in the following. We will show that if  there exists a bounded function $v(\bm{x})$ satisfying $\mathbb{E}^{\infty}[v(\bm{\phi}_{\pi}^{\bm{x}}(1))]-v(\bm{x})\leq 0, \forall \bm{x}\in \mathcal{X}$ and $0\leq v(\bm{x}), \forall \bm{x}\in \mathcal{X}$, there exists a bounded function $w(\bm{x}):\mathcal{X}\rightarrow \mathbb{R}$ such that 
\begin{equation*}
    \begin{cases}
         &v(\bm{x})\geq \mathbb{E}^{\infty}[v(\bm{\phi}^{\bm{x}}_{\pi}(1))], \forall \bm{x}\in \mathcal{X}\setminus \mathcal{X}_r,\\
       &v(\bm{x})\geq \mathbb{E}^{\infty}[w(\bm{\phi}^{\bm{x}}_{\pi}(1))]-w(\bm{x}), \forall \bm{x}\in \mathcal{X}\setminus \mathcal{X}_r.
    \end{cases}
\end{equation*}
Obviously, $w(\bm{x})\equiv 0$ for $\bm{x}\in \mathcal{X}$ satisfies this condition. Therefore, the conclusion holds and  constraint \eqref{sb1} is weaker than constraint \eqref{sbs}.

\begin{remark}
    When Assumption \ref{asump} does not hold, we have the following sufficient condition for certifying upper bounds of the reachability probability. Similar to Proposition \ref{reachability}, this condition can be obtained via relaxing \eqref{reach_equa}.
    \begin{proposition}
    \label{reachability_up}
If there exist bounded functions $v(\bm{x}):\mathcal{X}\rightarrow \mathbb{R}$ and $w(\bm{x}): \mathcal{X}\rightarrow \mathbb{R}$ such that 
 \begin{equation*}
   \begin{cases}
       &v(\bm{x}) \leq \epsilon_2, \forall \bm{x}\in \mathcal{X}_0,\\
       &v(\bm{x})\geq \mathbb{E}^{\infty}[v(\widehat{\bm{\phi}}^{\bm{x}}_{\pi}(1))], \forall \bm{x}\in \widehat{\mathcal{X}},\\
       &v(\bm{x})\geq 1_{\mathcal{X}_r}(\bm{x})+\mathbb{E}^{\infty}[w(\widehat{\bm{\phi}}^{\bm{x}}_{\pi}(1))]-w(\bm{x}), \forall \bm{x}\in \widehat{\mathcal{X}},
   \end{cases}
   \end{equation*}
   which is equivalent to 
   \begin{equation}
   \begin{cases}
      &v(\bm{x})\leq \epsilon_2,\forall \bm{x}\in \mathcal{X}_0,\\ 
       &v(\bm{x})\geq \mathbb{E}^{\infty}[v(\bm{\phi}^{\bm{x}}_{\pi}(1))], \forall \bm{x}\in \mathcal{X}\setminus \mathcal{X}_r,\\
       &v(\bm{x})\geq \mathbb{E}^{\infty}[w(\bm{\phi}^{\bm{x}}_{\pi}(1))]-w(\bm{x}), \forall \bm{x}\in \mathcal{X}\setminus \mathcal{X}_r,\\
       &v(\bm{x})\geq 1, \forall \bm{x}\in \mathcal{X}_r,\\
      &v(\bm{x})\geq 0, \forall \bm{x}\in \widehat{\mathcal{X}}\setminus \mathcal{X},
   \end{cases}
   \end{equation}
   then $\mathbb{P}^{\infty}\big(\exists k\in \mathbb{N}. \bm{\phi}^{\bm{x}}_{\pi}(k) \in \mathcal{X}_r \wedge \forall i\in [0,k]. \bm{\phi}_{\pi}^{\bm{x}}(i)\in \mathcal{X} \mid  \bm{x}\in \mathcal{X}_0 \big)\leq \epsilon_2$.
    \end{proposition}
    \pf{
  The conclusion can be assured by following the proof of Corollary 2 in \cite{xue2021reach}.     \qed
 }      
\end{remark}

\begin{remark}
When Assumption \ref{asump} does not hold, inspired by \cite{xue2022reach}, we can further consider the case that $w(\bm{x})=\lambda v(\bm{x})$ for $\bm{x}\in \widehat{\mathcal{X}}$ in Proposition \ref{reachability} and \ref{reachability_up} for constructing sufficient conditions for reachability verification.

Via setting $w(\bm{x}):=\lambda v(\bm{x})$ and removing the constraint $v(\bm{x})\leq \mathbb{E}^{\infty}[v(\widehat{\bm{\phi}}^{\bm{x}}_{\pi}(1))], \forall \bm{x}\in \widehat{\mathcal{X}}$ in Proposition \ref{reachability}, we have the following  sufficient condition for certifying lower bounds of the reachability probability in Definition \ref{ravoid}. Its proof is shown in Appendix.
    \begin{proposition}
    \label{pro_e1}
    If there exist a bounded function $v(\bm{x}):\widehat{\mathcal{X}}\rightarrow \mathbb{R}$ and a positive value $\lambda \in (0,\infty)$ such that 
   \begin{equation}
   \label{ebs_low}
   \begin{cases}
   &v(\bm{x})\geq \epsilon_1,\forall \bm{x}\in \mathcal{X}_0,\\
   &v(\bm{x})\leq 1_{\mathcal{X}_r}(\bm{x})+\lambda(\mathbb{E}^{\infty}[v(\widehat{\bm{\phi}}_{\pi}^{\bm{x}}(1))]-v(\bm{x})), \forall \bm{x}\in \widehat{\mathcal{X}},
   \end{cases}
   \end{equation}
   which is equivalent to 
    \begin{equation*}
   \label{ebr_low}
   \begin{cases}
   &v(\bm{x})\geq \epsilon_1,\forall \bm{x}\in \mathcal{X}_0,\\
   &v(\bm{x})\leq \lambda(\mathbb{E}^{\infty}[v(\bm{\phi}_{\pi}^{\bm{x}}(1))]- v(\bm{x})), \forall \bm{x}\in \mathcal{X}\setminus \mathcal{X}_r,\\
   &v(\bm{x})\leq 1, \forall \bm{x}\in \mathcal{X}_r,\\
   &v(\bm{x})\leq 0, \forall \bm{x}\in  \widehat{\mathcal{X}}\setminus \mathcal{X},
   \end{cases}
   \end{equation*}
   then $\mathbb{P}^{\infty}\big(\exists k \in \mathbb{N}. \bm{\phi}^{\bm{x}_0}_{\pi}(k) \in \mathcal{X}_r\bigwedge \forall l\in [0,k]. \bm{\phi}^{\bm{x}_0}_{\pi}(l)\in \mathcal{X} \mid  \bm{x}_0\in \mathcal{X}_0 \big)= \mathbb{P}^{\infty}\big(\exists k\in \mathbb{N}. \widehat{\bm{\phi}}^{\bm{x}_0}_{\pi}(k) \in \mathcal{X}_r \mid  \bm{x}_0\in \mathcal{X}_0 \big)\geq \epsilon_1$. 
    \end{proposition}

  However, via setting $w(\bm{x}):=\lambda v(\bm{x})$ and removing $v(\bm{x})\geq \mathbb{E}^{\infty}[v(\widehat{\bm{\phi}}^{\bm{x}}_{\pi}(1))], \forall \bm{x}\in \widehat{\mathcal{X}}$ in Proposition \ref{reachability_up}, we cannot obtain sufficient conditions for certifying upper bounds of the reachability probability in Definition \ref{ravoid}. In contrast, we can obtain a sufficient condition for certifying upper bounds of the probability, with which the system starting from each state in $\mathcal{X}_0$ will enter the target set $\mathcal{X}_r$ within a uniformly unbounded time horizon.
    \begin{proposition}
    \label{pro_e2}
    If there exist a bounded function $v(\bm{x}):\widehat{\mathcal{X}}\rightarrow \mathbb{R}$ and a positive value $\lambda \in (0,\infty)$ such that 
   \begin{equation}
   \label{ebs}
   \begin{cases}
   &v(\bm{x})\leq \epsilon'_2,\forall \bm{x}\in \mathcal{X}_0,\\
   &v(\bm{x})\geq 1_{\mathcal{X}_r}(\bm{x})+\lambda(\mathbb{E}^{\infty}[v(\widehat{\bm{\phi}}_{\pi}^{\bm{x}}(1))]-v(\bm{x})), \forall \bm{x}\in \widehat{\mathcal{X}},
   \end{cases}
   \end{equation}
   which is equivalent to 
    \begin{equation*}
   \label{ebr}
   \begin{cases}
   &v(\bm{x})\leq \epsilon'_2,\forall \bm{x}\in \mathcal{X}_0,\\
   &v(\bm{x})\geq \lambda(\mathbb{E}^{\infty}[v(\bm{\phi}_{\pi}^{\bm{x}}(1))]- v(\bm{x})), \forall \bm{x}\in \mathcal{X}\setminus \mathcal{X}_r,\\
   &v(\bm{x})\geq 1, \forall \bm{x}\in \mathcal{X}_r,\\
   &v(\bm{x})\geq 0, \forall \bm{x}\in  \widehat{\mathcal{X}}\setminus \mathcal{X},
   \end{cases}
   \end{equation*}
   where $\epsilon'_2\in [0,1]$, then $\mathbb{P}^{\infty}\big(\exists k\in [0,N]. \bm{\phi}^{\bm{x}_0}_{\pi}(k) \in \mathcal{X}_r\bigwedge \forall l\in [0,k]. \bm{\phi}^{\bm{x}_0}_{\pi}(l)\in \mathcal{X} \mid  \bm{x}_0\in \mathcal{X}_0 \big)=\mathbb{P}^{\infty}\big(\exists k\in [0,N]. \widehat{\bm{\phi}}^{\bm{x}_0}_{\pi}(k) \in \mathcal{X}_r \mid  \bm{x}_0\in \mathcal{X}_0 \big)\leq \frac{(1+\lambda)^N}{\lambda^N}\epsilon'_2$. 
    \end{proposition}
    \pf{Its proof is shown in Appendix. \qed}
    
The result in Proposition \ref{pro_e2} complements the one in Proposition 2 in \cite{santoyo2021barrier} with $\widetilde{\alpha}<1$ and $\widetilde{\beta}=0$. However, the function in \eqref{ebr} is not required to be non-negative over $\widehat{\mathcal{X}}$. A general condition, which complements the one in Proposition 2 in \cite{santoyo2021barrier}, is formulated below.  It requires the function $v(\bm{x})$ to be non-negative over $\widehat{\mathcal{X}}$. 
\begin{proposition}
    \label{pro_e4}
    If there exist a function $v(\bm{x}):\widehat{\mathcal{X}}\rightarrow \mathbb{R}$, and $\widetilde{\alpha}\in (0,1]$ and $0\leq \widetilde{\beta}<1$ such that 
   \begin{equation}
   \label{ebs_3}
   \begin{cases}
   &v(\bm{x})\leq \epsilon'_2,\forall \bm{x}\in \mathcal{X}_0,\\
   &v(\bm{x})\geq \widetilde{\alpha} \mathbb{E}^{\infty}[v(\widehat{\bm{\phi}}_{\pi}^{\bm{x}}(1))]-\widetilde{\alpha} \widetilde{\beta}, \forall \bm{x}\in \widehat{\mathcal{X}},\\
   &v(\bm{x})\geq 1_{\mathcal{X}_r}(\bm{x}), \forall \bm{x}\in \widehat{\mathcal{X}},
   \end{cases}
   \end{equation}
   which is equivalent to 
    \begin{equation}
   \label{ebr_4}
   \begin{cases}
   &v(\bm{x})\leq \epsilon'_2,\forall \bm{x}\in \mathcal{X}_0,\\
   &v(\bm{x})\geq \widetilde{\alpha} \mathbb{E}^{\infty}[v(\bm{\phi}_{\pi}^{\bm{x}}(1))]-\widetilde{\alpha} \widetilde{\beta}, \forall \bm{x}\in \mathcal{X}\setminus \mathcal{X}_r,\\
   &v(\bm{x})\geq 1, \forall \bm{x}\in \mathcal{X}_r,\\
   &v(\bm{x})\geq 0, \forall \bm{x}\in  \mathcal{X}\setminus \mathcal{X}_r,\\
      &v(\bm{x})\geq 0, \forall \bm{x}\in  \widehat{\mathcal{X}}\setminus \mathcal{X}, 
   \end{cases}
   \end{equation}
   where $\epsilon'_2\in [0,1]$, then $\mathbb{P}^{\infty}\big(\exists k\in [0,N]. \bm{\phi}^{\bm{x}_0}_{\pi}(k) \in \mathcal{X}_r\bigwedge \forall l\in [0,k]. \bm{\phi}^{\bm{x}_0}_{\pi}(l)\in \mathcal{X} \mid  \bm{x}_0\in \mathcal{X}_0 \big)=\mathbb{P}^{\infty}\big(\exists k\in [0,N]. \widehat{\bm{\phi}}^{\bm{x}_0}_{\pi}(k) \in \mathcal{X}_r \mid  \bm{x}_0\in \mathcal{X}_0 \big)= P$, where 
   \begin{enumerate}
       \item if $\widetilde{\alpha}=1$,
       $P\leq \epsilon'_2+\widetilde{\beta}N$.
       \item if $\widetilde{\alpha}<1$, $P\leq \epsilon'_2\widetilde{\alpha}^{-N}+\frac{(1-\widetilde{\alpha}^{-N})\widetilde{\alpha}\widetilde{\beta}}{\widetilde{\alpha}-1}$.
   \end{enumerate}
    \end{proposition}
    \pf{Its proof is shown in Appendix. \qed}
    
   When $\widetilde{\alpha}=\frac{\lambda}{1+\lambda}$ and $\widetilde{\beta}=0$ in Proposition \ref{pro_e2}, we can obtain the conclusion in Proposition \ref{pro_e4}. Under Assumption \ref{asump}, the constraint $v(\bm{x})\geq 0, \forall \bm{x}\in  \widehat{\mathcal{X}}\setminus \mathcal{X}$ in \eqref{ebr_4} is redundant and thus can be removed. A weaker condition, which does not require $v(\bm{x})$ to be non-negative over $\widehat{\mathcal{X}}$, is presented below.
\begin{proposition}
     \label{pro_e3}
    If there exist a bounded function $v(\bm{x}):\widehat{\mathcal{X}}\rightarrow \mathbb{R}$, and positive values $\widetilde{\alpha} \in (0,1)$ and $\widetilde{\beta} \in [0,1)$, such that 
   \begin{equation}
   \label{ebs1}
   \begin{cases}
   &v(\bm{x})\leq \epsilon'_2,\forall \bm{x}\in \mathcal{X}_0,\\
   &v(\bm{x})\geq (1+\widetilde{\alpha}\widetilde{\beta}-\widetilde{\alpha})1_{\mathcal{X}_r}(\bm{x})\\
   &~~~~~~~~~~~~~~~~+\widetilde{\alpha}\mathbb{E}^{\infty}[v(\widehat{\bm{\phi}}_{\pi}^{\bm{x}}(1))]-\widetilde{\alpha}\widetilde{\beta}, \forall \bm{x}\in \widehat{\mathcal{X}},
   \end{cases}
   \end{equation}
   which is equivalent to 
    \begin{equation}
   \label{ebr1}
   \begin{cases}
   &v(\bm{x})\leq \epsilon'_2,\forall \bm{x}\in \mathcal{X}_0,\\
   &v(\bm{x})\geq \widetilde{\alpha}\mathbb{E}^{\infty}[v(\bm{\phi}_{\pi}^{\bm{x}}(1))]-\widetilde{\alpha}\widetilde{\beta}, \forall \bm{x}\in \mathcal{X}\setminus \mathcal{X}_r,\\
   &v(\bm{x})\geq 1, \forall \bm{x}\in \mathcal{X}_r,\\
   &v(\bm{x})\geq -\frac{\widetilde{\alpha}\widetilde{\beta}}{1-\widetilde{\alpha}}, \forall \bm{x}\in  \widehat{\mathcal{X}}\setminus \mathcal{X},
   \end{cases}
   \end{equation}
   where $\epsilon'_2\in [0,1]$, then $\mathbb{P}^{\infty}\big(\exists k\in [0,N]. \bm{\phi}^{\bm{x}_0}_{\pi}(k) \in \mathcal{X}_r\bigwedge \forall l\in [0,k]. \bm{\phi}^{\bm{x}_0}_{\pi}(l)\in \mathcal{X} \mid  \bm{x}_0\in \mathcal{X} \big)=\mathbb{P}^{\infty}\big(\exists k\in [0,N]. \widehat{\bm{\phi}}^{\bm{x}_0}_{\pi}(k) \in \mathcal{X}_r \mid  \bm{x}_0\in \mathcal{X}_0 \big)\leq \frac{\epsilon'_2\widetilde{\alpha}^{-N}(1-\widetilde{\alpha})+\widetilde{\alpha}\widetilde{\beta}\widetilde{\alpha}^{-N}}{1+\widetilde{\alpha}\widetilde{\beta}-\widetilde{\alpha}}$. 
\end{proposition}

If a function $v(\bm{x})$ satisfies \eqref{ebr_4} with $\widetilde{\alpha}\in (0,1)$, it also satisfies \eqref{ebr1}. Also, when $\widetilde{\alpha}^N\leq \epsilon'_2$, $\frac{\epsilon'_2\widetilde{\alpha}^{-N}(1-\widetilde{\alpha})+\widetilde{\alpha}\widetilde{\beta}\widetilde{\alpha}^{-N}}{1+\widetilde{\alpha}\widetilde{\beta}-\widetilde{\alpha}}\leq \epsilon'_2\widetilde{\alpha}^{-N}+\frac{(1-\widetilde{\alpha}^{-N})\widetilde{\alpha}\widetilde{\beta}}{\widetilde{\alpha}-1}$ holds. \qed
\end{remark}

\section{$k$-Inductive Reachability Verification}
\label{k_R_V}
In this section we extend conditions in Proposition \ref{reachability} and \ref{SB} based on the $k$-induction principle, where $k\in \mathbb{N}$, and present new sufficient conditions for reachability verification in Definition \ref{ravoid}.

We first present the sufficient condition for certifying lower bounds of the reachability probability. This condition is an extension of Proposition \ref{reachability} to the $k$-composition $\bm{x}(l+1)=\bm{f}^k(\bm{x}(l),\bm{\theta}(kl:{kl+k-1}))$ of system \eqref{system}. 

\begin{proposition}
\label{k_reach}
Under Assumption \ref{asump}, if there exist bounded functions $v(\bm{x}):\mathcal{X}\rightarrow \mathbb{R}$ and $w(\bm{x}): \mathcal{X}\rightarrow \mathbb{R}$ such that 
\begin{equation}
\label{k_r_b11}
   \begin{cases}
       &v(\bm{x})\geq 1-\epsilon'_1, \forall \bm{x}\in \mathcal{X}_0,\\
       &  v(\bm{x})\leq \mathbb{E}^{\infty}[v(\bm{\phi}_{\pi}^{\bm{x}}(k))], \forall \bm{x}\in \mathcal{X}\setminus \mathcal{X}_r,\\
       &v(\bm{x})\leq \mathbb{E}^{\infty}[w(\bm{\phi}_{\pi}^{\bm{x}}(k))]-w(\bm{x}), \forall \bm{x}\in \mathcal{X}\setminus \mathcal{X}_r,\\
       & v(\bm{x})\leq 1, \forall \bm{x}\in \mathcal{X}_r,
   \end{cases}
   \end{equation}
   where $\epsilon'_1\in [0,1]$,
      then $\mathbb{P}^{\infty}\big(\exists i\in \mathbb{N}. \bm{\phi}^{\bm{x}}_{\pi}(i) \in \mathcal{X}_r \mid \bm{x}\in \mathcal{X}_0\big)\geq 1-\epsilon'_1$.
\end{proposition}
\pf{
From constraint \eqref{k_r_b11}, we have that 
\begin{equation*}
\begin{cases}
    &v(\bm{x})\geq 1-\epsilon'_1, \forall \bm{x}\in \mathcal{X}_0,\\
    &v(\bm{x})\leq \mathbb{E}^{\infty}[v(\check{\bm{\phi}}_{\pi}^{\bm{x}}(1))], \forall \bm{x}\in \mathcal{X},\\
    &v(\bm{x})\leq 1_{\mathcal{X}_r}(\bm{x})+\mathbb{E}^{\infty}[w(\check{\bm{\phi}}_{\pi}^{\bm{x}}(1))]-w(\bm{x}), \forall \bm{x}\in \mathcal{X},
\end{cases}
\end{equation*}
where $\check{\bm{\phi}}_{\pi}^{\bm{x}}(\cdot):\mathbb{N}\rightarrow \mathbb{R}^n$ satisfies:
\begin{equation*}
    \label{new_system2}
\begin{cases}
&\check{\bm{\phi}}_{\pi}^{\bm{x}}(j+1)=1_{\mathcal{X}_r}(\check{\bm{\phi}}_{\pi}^{\bm{x}}(j))\cdot \check{\bm{\phi}}_{\pi}^{\bm{x}}(j)+\\
&1_{\mathcal{X}\setminus \mathcal{X}_r}(\check{\bm{\phi}}_{\pi}^{\bm{x}}(j))\cdot \bm{f}^k(\check{\bm{\phi}}_{\pi}^{\bm{x}}(j),\bm{\theta}(jk:jk+k-1)), \forall j\in \mathbb{N},\\
&\check{\bm{\phi}}_{\pi}^{\bm{x}}(0)=\bm{x}.
\end{cases}
\end{equation*}

Therefore, we have that for $\bm{x}\in \mathcal{X}$,
\begin{equation*}
\begin{split}
    &v(\bm{x})\leq 1_{\mathcal{X}_r}(\bm{x})+\mathbb{E}^{\infty}[w(\check{\bm{\phi}}_{\pi}^{\bm{x}}(1))]-w(\bm{x}),\\
    &v(\bm{x})\leq \mathbb{E}^{\infty}[v(\check{\bm{\phi}}_{\pi}^{\bm{x}}(1))]\leq \mathbb{E}^{\infty}[1_{\mathcal{X}_r}(\check{\bm{\phi}}_{\pi}^{\bm{x}}(1))]\\
    &~~~~~~~~~~~~~~~~~~~~~~~~~~~~+\mathbb{E}^{\infty}[w(\check{\bm{\phi}}_{\pi}^{\bm{x}}(2))]-\mathbb{E}^{\infty}[w(\check{\bm{\phi}}_{\pi}^{\bm{x}}(1)],\\
    &\ldots,\\
    &v(\bm{x})\leq \mathbb{E}^{\infty}[v(\check{\bm{\phi}}_{\pi}^{\bm{x}}(m-1))]\leq \mathbb{E}^{\infty}[1_{\mathcal{X}_r}(\check{\bm{\phi}}_{\pi}^{\bm{x}}(m-1))]\\
    &~~~~~~~~~~~~~+\mathbb{E}^{\infty}[w(\check{\bm{\phi}}_{\pi}^{\bm{x}}(m))]-\mathbb{E}^{\infty}[w(\check{\bm{\phi}}_{\pi}^{\bm{x}}(m-1))],
\end{split}
\end{equation*}
and thus 
\[
\begin{split}
&m v(\bm{x})\leq 1_{\mathcal{X}_r}(\bm{x})+ \mathbb{E}^{\infty}[1_{\mathcal{X}_r}(\check{\bm{\phi}}_{\pi}^{\bm{x}}(1))]+\ldots \\
&+\mathbb{E}^{\infty}[1_{\mathcal{X}_r}(\check{\bm{\phi}}_{\pi}^{\bm{x}}(m-1))]
+\mathbb{E}^{\infty}[w(\check{\bm{\phi}}_{\pi}^{\bm{x}}(m))]-w(\bm{x}).
\end{split}\]
Consequently, as $m\rightarrow \infty$, together with $v(\bm{x})\geq 1-\epsilon’_1, \forall \bm{x}\in \mathcal{X}_0$,  we have 
\[
\begin{split}
(1-\epsilon'_1) 
&\leq \lim_{m\rightarrow \infty} \frac{1_{\mathcal{X}_r}(\bm{x})+\ldots +\mathbb{E}^{\infty}[1_{\mathcal{X}_r}(\check{\bm{\phi}}_{\pi}^{\bm{x}}(m-1))]}{m}\\
&=\mathbb{P}^{\infty}\big( \exists j\in \mathbb{N}. \bm{\phi}_{\pi}^{\bm{x}}(jk)\in \mathcal{X}_r\mid \bm{x}\in \mathcal{X}_0\big)\\
&\leq \mathbb{P}^{\infty}\big( \exists i\in \mathbb{N}. \bm{\phi}_{\pi}^{\bm{x}}(i)\in \mathcal{X}_r\mid \bm{x}\in \mathcal{X}_0\big).
\end{split}
\]
The proof is completed. \qed
}

From \eqref{k_r_b11}, we can observe that for any state $\bm{x}$ in $\{\bm{x}\in  \mathcal{X} \mid  v(\bm{x}) \geq 1-\epsilon'_1\}$, both the $k$-composition system $\bm{x}(l+1)=\bm{f}^k(\bm{x}(l),\bm{\theta}({kl}:{kl+k-1}))$ and system \eqref{system} starting from it will enter the target set $\mathcal{X}_r$ with a probability being larger than or equal to  $1-\epsilon'_1$. Therefore, by setting $\epsilon'_1:=1-\epsilon_1$ in Proposition \ref{k_reach}, we have that $\mathbb{P}^{\infty}\big( \exists i\in \mathbb{N}. \bm{\phi}_{\pi}^{\bm{x}}(i)\in \mathcal{X}_r\mid \bm{x}\in \mathcal{X}_0\big)\geq \epsilon_1$ if constraint \eqref{k_r_b11} holds. 

Via setting $u(\bm{x}):=1-v(\bm{x})$ in Proposition \ref{k-inductive_reachability}, constraint \eqref{k-inductive_reachability1} can be equivalently reformulated as 
\begin{equation}
\label{k_r_b110}
   \begin{cases}
       &u(\bm{x})\geq 1-\epsilon'_1, \forall \bm{x}\in \mathcal{X}_0,\\
       &  u(\bm{x})\leq 0, \forall \bm{x}\in \partial \mathcal{X}\setminus \partial \mathcal{X}_r,\\
      &\mathbb{E}^{\infty}[u(\bm{\phi}_{\pi}^{\bm{x}}(1))]-u(\bm{x})\geq -c, \forall \bm{x}\in \overline{\mathcal{X}\setminus \mathcal{X}_r},\\
            &\mathbb{E}^{\infty}[u(\bm{\phi}_{\pi}^{\bm{x}}(k))]-u(\bm{x})\geq \delta, \forall \bm{x}\in \overline{\mathcal{X}\setminus \mathcal{X}_r},\\
       & u(\bm{x})\leq 1, \forall \bm{x}\in \mathcal{X}.
   \end{cases}
   \end{equation}
 Using the same inference technique of comparing constraints \eqref{super_reach} and \eqref{rb1} in Section \ref{sec:comp}, we can conclude that constraint \eqref{k_r_b11} is weaker than \eqref{k-inductive_reachability1}. Moreover, comparing the lower bounds (i.e., $1-k\epsilon'_1-\frac{k(k-1)}{2}c$ and $1-\epsilon'_1$) in constraints \eqref{k-inductive_reachability1} 
 and \eqref{k_r_b11}, we can also have the conclusion that constraint \eqref{k_r_b11} is able to certify tighter lower bounds of the reachability probability than \eqref{k-inductive_reachability1}.

Next, we present a sufficient condition of certifying upper bounds of the reachability probability for system \eqref{system}. Similar to Proposition \ref{k_reach}, this sufficient condition is an extension of Proposition \ref{SB} to the $k$-fold system $\bm{x}(l+1)=\bm{f}^k(\bm{x}(l),\theta(kl:{kl+k-1}))$ starting from the set $\mathcal{X}_0$.   

\begin{proposition}
\label{k_reach1}
Under Assumption \ref{asump}, if there exist bounded functions $v(\bm{x}):\widehat{\mathcal{X}}\rightarrow \mathbb{R}$ and $w(\bm{x}): \widehat{\mathcal{X}}\rightarrow \mathbb{R}$ such that 
\begin{equation}
\label{k_r_b111}
   \begin{cases}
       &v(\bm{x})\leq \epsilon'_2, \forall \bm{x}\in \mathcal{X}_0,\\
       &c+ v(\bm{x})\geq \mathbb{E}^{\infty}[v(\bm{\phi}_{\pi}^{\bm{x}}(1))], \forall \bm{x}\in \mathcal{X},\\
       &  v(\bm{x})\geq \mathbb{E}^{\infty}[v(\bm{\phi}_{\pi}^{\bm{x}}(k))], \forall \bm{x}\in \mathcal{X}\setminus \mathcal{X}_r,\\
       &v(\bm{x})\geq \mathbb{E}^{\infty}[w(\bm{\phi}_{\pi}^{\bm{x}}(k))]-w(\bm{x}), \forall \bm{x}\in \mathcal{X}\setminus \mathcal{X}_r,\\
       &v(\bm{x})\geq 1, \forall \bm{x}\in \mathcal{X}_r,
   \end{cases}
   \end{equation}
   where $c\geq 0$ is a user-defined value and $\epsilon'_2\in [0,1]$, then $\mathbb{P}^{\infty}\big(\exists i\in \mathbb{N}. \bm{\phi}^{\bm{x}}_{\pi}(i) \in \mathcal{X}_r \mid \bm{x}\in \mathcal{X}_0\big)\leq k\epsilon'_2+\frac{k(k-1)c}{2}$. 
\end{proposition}
\pf{Following the proof of Proposition \ref{k_reach} we have that 
\[\mathbb{P}^{\infty}\big(\exists j\in \mathbb{N}. \bm{\phi}^{\bm{x}}_{\pi}(jk) \in \mathcal{X}_r \mid \bm{x}\in \mathcal{X}\big)\leq v(\bm{x}).\]

Let \[A=\{\pi\mid \exists i\in \mathbb{N}. \bm{\phi}^{\bm{x}}_{\pi}(i) \in \mathcal{X}_r \mid \bm{x}\in \mathcal{X}\}\] and 
\[A_i=\{\pi\mid \exists j\in \mathbb{N}. \bm{\phi}^{\bm{x}}_{\pi}(jk+i) \in \mathcal{X}_r \mid \bm{x}\in \mathcal{X}\}.\]
Thus, $A=\cup_{i=0}^{k-1}A_i$ and $\mathbb{P}^{\infty}(A)\leq \sum_{i=0}^{k-1} \mathbb{P}^{\infty}(A_i)$.

Taking $\bm{x}_i=\bm{\phi}^{\bm{x}}_{\pi}(i)$ for $i=1,\ldots,k-1$, we have that 
\[\mathbb{P}^{\infty}\big(\exists j\in \mathbb{N}. \bm{\phi}^{\bm{x}_i}_{\pi}(jk) \in \mathcal{X}_r \mid \bm{x}_i\in \mathcal{X}\big)\leq v(\bm{x}_i),\]
which implies that 
\[\begin{split}
&\mathbb{P}^{\infty}\big(\exists j\in \mathbb{N}. \bm{\phi}^{\bm{x}}_{\pi}(jk+i) \in \mathcal{X}_r \mid \bm{x}\in \mathcal{X}\big)\\
&=\mathbb{P}^{\infty}(A_i)\leq \mathbb{E}^{\infty}[v(\bm{\phi}^{\bm{x}}_{\pi}(i))].
\end{split}\]
Since $v(\bm{x})+c\geq \mathbb{E}^{\infty}[v(\bm{\phi}_{\pi}^{\bm{x}}(1))], \forall \bm{x}\in \mathcal{X}$, we have that \[\mathbb{E}^{\infty}[v(\bm{\phi}^{\bm{x}}_{\pi}(i))]\leq v(\bm{x})+ic\]
for $i=1,\ldots,k-1$.

Also, since  $v(\bm{x})\leq \epsilon'_2, \forall \bm{x}\in \mathcal{X}_0$ and $\mathbb{P}^{\infty}(A)\leq \sum_{i=0}^{k-1} \mathbb{P}^{\infty}(A_i)$, we can obtain 
\[\mathbb{P}^{\infty}\big(\exists i\in \mathbb{N}. \bm{\phi}^{\bm{x}}_{\pi}(i) \in \mathcal{X}_r \mid \bm{x}\in \mathcal{X}_0\big)\leq k\epsilon_2'+\frac{k(k-1)c}{2}.\]
The proof is completed.
\qed}

By setting $\epsilon'_2:=\frac{2\epsilon_2-k(k-1)c}{2k}$ in Proposition \ref{k_reach1}, we have $\mathbb{P}^{\infty}\big( \exists i\in \mathbb{N}. \bm{\phi}_{\pi}^{\bm{x}}(i)\in \mathcal{X}_r\mid \bm{x}\in \mathcal{X}_0\big)\leq \epsilon_2$ if constraint \eqref{k_r_b111} holds. Analogously, we can conclude that constraint \eqref{k_r_b111} is weaker, comparing to constraint \eqref{k-inductive_safe1}.

\begin{remark}
In the set of constraints \eqref{k_r_b111},  the constraint 
\[c+ v(\bm{x})\geq \mathbb{E}^{\infty}[v(\bm{\phi}_{\pi}^{\bm{x}}(1))], \forall \bm{x}\in \mathcal{X}\] can be  replaced with
\[\mathbb{E}^{\infty}[v(\bm{\phi}_{\pi}^{\bm{x}}(1))]-\alpha v(\bm{x})\leq 0, \forall \bm{x}\in \mathcal{X},\] where $\alpha \in (0,\infty)$. Then, we will have the conclusion that 
\[\mathbb{P}^{\infty}\big(\exists i\in \mathbb{N}. \bm{\phi}^{\bm{x}}_{\pi}(i) \in \mathcal{X}_r \mid \bm{x}\in \mathcal{X}_0\big)\leq \epsilon'_2 \frac{1-\alpha^k}{(1-\alpha)}\] in Proposition \ref{k_reach1}. \qed
\end{remark}

\begin{remark}
When $c=0$ in \eqref{k_r_b111}, the constraint $v(\bm{x})\geq \mathbb{E}^{\infty}[v(\bm{\phi}^{\bm{x}}_{\pi}(k))], \forall \bm{x}\in \mathcal{X}\setminus \mathcal{X}_r$ is redundant and thus can be removed from \eqref{k_r_b111}. Furthermore, it is interesting to find that when $c=0$ and $w(\bm{x})=0$ for $\bm{x}\in \mathcal{X}$, a function $v(\bm{x})$ satisfying \eqref{k_r_b111} also satisfies \eqref{sbs}, but we obtain a conservative conclusion from Proposition \ref{k_reach1}, which is 
\[\mathbb{P}^{\infty}\big(\exists i\in \mathbb{N}. \bm{\phi}^{\bm{x}}_{\pi}(i) \in \mathcal{X}_r\mid \bm{x}\in \mathcal{X}_0\big)\leq k\epsilon'_2\]
rather than \[\mathbb{P}^{\infty}\big(\exists i\in \mathbb{N}. \bm{\phi}^{\bm{x}}_{\pi}(i) \in \mathcal{X}_r\mid \bm{x}\in \mathcal{X}_0\big)\leq \epsilon'_2.\] The following condition will remedy this issue using system \eqref{new_system1}.
\begin{proposition}
\label{upper_invariance1}
Under Assumption \ref{asump}, if there exist bounded functions $v(\bm{x}): \mathcal{X}\rightarrow \mathbb{R}$ and $w(\bm{x}): \mathcal{X}\rightarrow \mathbb{R}$ satisfying 
   \begin{equation}
   \label{reach_equa4}
   \begin{cases}
       &v(\bm{x})\leq \epsilon'_2, \forall \bm{x}\in \mathcal{X}_0,\\
       &v(\bm{x})\geq \mathbb{E}^{\infty}[v(\bm{\phi}^{\bm{x}}_{\pi}(1))]-c, \forall \bm{x}\in \mathcal{X}\setminus \mathcal{X}_r,\\
       &v(\bm{x})\geq \mathbb{E}^{\infty}[v(\widetilde{\bm{\phi}}^{\bm{x}}_{\pi}(k))], \forall \bm{x}\in \mathcal{X},\\
       &v(\bm{x})\geq \mathbb{E}^{\infty}[w(\bm{\phi}^{\bm{x}}_{\pi}(1))]-w(\bm{x}), \forall \bm{x}\in \mathcal{X}\setminus \mathcal{X}_r,\\
       &v(\bm{x})\geq 1, \forall \bm{x}\in \mathcal{X}_r,
   \end{cases}
   \end{equation}
  where $c\geq 0$ is a user-defined value, $\epsilon'_2\in [0,1]$ and $\widetilde{\bm{\phi}}^{\bm{x}}_{\pi}(\cdot): \mathbb{N}\rightarrow \mathcal{X}$ is the trajectory to system \eqref{new_system1}, then $\mathbb{P}^{\infty}\big(\exists i\in \mathbb{N}. \bm{\phi}^{\bm{x}}_{\pi}(i) \in \mathcal{X}_r\mid \bm{x}\in \mathcal{X}_0\big)\leq \epsilon'_2+\frac{(k-1)c}{2}$.
\end{proposition}  
Its proof is presented in Appendix. \qed
\end{remark}

\section{Conclusion}
\label{sec:con}
In this paper we presented new sufficient conditions for reachability verification over the infinite time horizon for stochastic discrete-time dynamical systems based on relaxing the set of equations in \cite{xue2021reach}. These sufficient conditions were shown to be weaker and more powerful in reachability verification than existing ones.   
\bibliographystyle{abbrv}
\bibliography{reference}

\begin{thebibliography}{10}

\bibitem{ames2019control}
A.~D. Ames, S.~Coogan, M.~Egerstedt, G.~Notomista, K.~Sreenath, and P.~Tabuada.
\newblock Control barrier functions: Theory and applications.
\newblock In {\em 2019 18th European control conference (ECC)}, pages
  3420--3431. IEEE, 2019.

\bibitem{anand2022k}
M.~Anand, V.~Murali, A.~Trivedi, and M.~Zamani.
\newblock k-inductive barrier certificates for stochastic systems.
\newblock In {\em 25th ACM International Conference on Hybrid Systems:
  Computation and Control}, pages 1--11, 2022.

\bibitem{baier2008principles}
C.~Baier and J.-P. Katoen.
\newblock {\em Principles of model checking}.
\newblock MIT press, 2008.

\bibitem{chakarov2013probabilistic}
A.~Chakarov and S.~Sankaranarayanan.
\newblock Probabilistic program analysis with martingales.
\newblock In {\em Computer Aided Verification: 25th International Conference,
  CAV 2013, Saint Petersburg, Russia, July 13-19, 2013. Proceedings 25}, pages
  511--526. Springer, 2013.

\bibitem{chatterjee2008model}
K.~Chatterjee, K.~Sen, and T.~A. Henzinger.
\newblock Model-checking $\omega$-regular properties of interval markov chains.
\newblock In {\em International Conference on Foundations of Software Science
  and Computational Structures}, pages 302--317. Springer, 2008.

\bibitem{franzle2008stochastic}
M.~Fr{\"a}nzle, H.~Hermanns, and T.~Teige.
\newblock Stochastic satisfiability modulo theory: A novel technique for the
  analysis of probabilistic hybrid systems.
\newblock In {\em International Workshop on Hybrid Systems: Computation and
  Control}, pages 172--186. Springer, 2008.

\bibitem{khalil2002nonlinear}
H.~K. Khalil.
\newblock Nonlinear systems third edition.
\newblock {\em Patience Hall}, 115, 2002.

\bibitem{lahijanian2015formal}
M.~Lahijanian, S.~B. Andersson, and C.~Belta.
\newblock Formal verification and synthesis for discrete-time stochastic
  systems.
\newblock {\em IEEE Transactions on Automatic Control}, 60(8):2031--2045, 2015.

\bibitem{lavaei2022automated}
A.~Lavaei, S.~Soudjani, A.~Abate, and M.~Zamani.
\newblock Automated verification and synthesis of stochastic hybrid systems: A
  survey.
\newblock {\em Automatica}, 146:110617, 2022.

\bibitem{oksendal2013stochastic}
B.~Oksendal.
\newblock {\em Stochastic differential equations: an introduction with
  applications}.
\newblock Springer Science \& Business Media, 2013.

\bibitem{prajna2007framework}
S.~Prajna, A.~Jadbabaie, and G.~J. Pappas.
\newblock A framework for worst-case and stochastic safety verification using
  barrier certificates.
\newblock {\em IEEE Transactions on Automatic Control}, 52(8):1415--1428, 2007.

\bibitem{prajna2007convex}
S.~Prajna and A.~Rantzer.
\newblock Convex programs for temporal verification of nonlinear dynamical
  systems.
\newblock {\em SIAM Journal on Control and Optimization}, 46(3):999--1021,
  2007.

\bibitem{santoyo2021barrier}
C.~Santoyo, M.~Dutreix, and S.~Coogan.
\newblock A barrier function approach to finite-time stochastic system
  verification and control.
\newblock {\em Automatica}, 125:109439, 2021.

\bibitem{sheeran2000checking}
M.~Sheeran, S.~Singh, and G.~St{\aa}lmarck.
\newblock Checking safety properties using induction and a sat-solver.
\newblock In {\em International conference on formal methods in computer-aided
  design}, pages 127--144. Springer, 2000.

\bibitem{steinhardt2012finite}
J.~Steinhardt and R.~Tedrake.
\newblock Finite-time regional verification of stochastic non-linear systems.
\newblock {\em The International Journal of Robotics Research}, 31(7):901--923,
  2012.

\bibitem{tkachev2011infinite}
I.~Tkachev and A.~Abate.
\newblock On infinite-horizon probabilistic properties and stochastic
  bisimulation functions.
\newblock In {\em 2011 50th IEEE Conference on Decision and Control and
  European Control Conference}, pages 526--531. IEEE, 2011.

\bibitem{tkachev2014characterization}
I.~Tkachev and A.~Abate.
\newblock Characterization and computation of infinite-horizon specifications
  over markov processes.
\newblock {\em Theoretical Computer Science}, 515:1--18, 2014.

\bibitem{xue2021reach}
B.~Xue, R.~Li, N.~Zhan, and M.~Fr{\"a}nzle.
\newblock Reach-avoid analysis for stochastic discrete-time systems.
\newblock In {\em 2021 American Control Conference (ACC)}, pages 4879--4885.
  IEEE, 2021.

\bibitem{xue2022reach1}
B.~Xue, N.~Zhan, and M.~Fr{\"a}nzle.
\newblock Reach-avoid analysis for stochastic differential equations.
\newblock {\em arXiv preprint arXiv:2208.10752}, 2022.

\bibitem{xue2022reach}
B.~Xue, N.~Zhan, M.~Fr{\"a}nzle, J.~Wang, and W.~Liu.
\newblock Reach-avoid verification based on convex optimization.
\newblock {\em arXiv preprint arXiv:2208.08105}, 2022.

\end{thebibliography}
\section*{Appendix}

\textbf{The proof of Proposition \ref{pro_e1}:}

\pf{
From \eqref{ebs_low}, we have that
\begin{equation*}
    \begin{split}
    &v(\bm{x})\leq \frac{1}{1+\lambda}1_{\mathcal{X}_r}(\bm{x})+\frac{\lambda}{1+\lambda}\mathbb{E}^{\infty}[v(\widehat{\bm{\phi}}_{\pi}^{\bm{x}}(1))],\\
    &\frac{\lambda}{1+\lambda} \mathbb{E}^{\infty}[v(\widehat{\bm{\phi}}_{\pi}^{\bm{x}}(1))]\leq \frac{\lambda}{(1+\lambda)^2}\mathbb{E}^{\infty}[1_{\mathcal{X}_r}(\widehat{\bm{\phi}}_{\pi}^{\bm{x}}(1))]\\
    &~~~~~~~~~~~~~~~~~~~~~~~~~~~~~~~~~~~~~~~~~~~~+\frac{\lambda^2}{(1+\lambda)^2}\mathbb{E}^{\infty}[v(\widehat{\bm{\phi}}_{\pi}^{\bm{x}}(2))],\\
    &\ldots,\\
    &\frac{\lambda^m}{(1+\lambda)^m} \mathbb{E}^{\infty}[v(\widehat{\bm{\phi}}_{\pi}^{\bm{x}}(m))]\leq \frac{\lambda^m}{(1+\lambda)^{m+1}} \mathbb{E}^{\infty}[1_{\mathcal{X}_r}(\widehat{\bm{\phi}}_{\pi}^{\bm{x}}(m))]\\
    &~~~~~~~~~~~~~~~~~~~~~~~~~~~~~~~~~+\frac{\lambda^{m+1}}{(1+\lambda)^{m+1}}\mathbb{E}^{\infty}[v(\widehat{\bm{\phi}}_{\pi}^{\bm{x}}(m+1))],\\
    &\ldots.
    \end{split}
\end{equation*}
Since $\mathbb{E}^{\infty}[1_{\mathcal{X}_r}(\widehat{\bm{\phi}}_{\pi}^{\bm{x}}(k))]=\mathbb{P}^{\infty}\big(\exists i\in [0,k]. \widehat{\bm{\phi}}^{\bm{x}}_{\pi}(i) \in \mathcal{X}_r\big), \forall k\in \mathbb{N}$ (this can be assured according to the fact that if $\widehat{\bm{\phi}}_{\pi}^{\bm{x}}(i)\in \mathcal{X}_r$, $\widehat{\bm{\phi}}_{\pi}^{\bm{x}}(j)\in \mathcal{X}_r$ for $j\geq i$), and $\mathbb{E}^{\infty}[1_{\mathcal{X}_r}(\widehat{\bm{\phi}}_{\pi}^{\bm{x}}(i))]\geq \mathbb{E}^{\infty}[1_{\mathcal{X}_r}(\widehat{\bm{\phi}}_{\pi}^{\bm{x}}(j))]$ for $i\geq j$, we can obtain that for $m\in \mathbb{N}$, 
\[
\begin{split}
&v(\bm{x})\leq \frac{\lambda^{m+1}}{(1+\lambda)^{m+1}}\mathbb{E}^{\infty}[v(\widehat{\bm{\phi}}_{\pi}^{\bm{x}}(m+1))]+\\
&\frac{1}{1+\lambda} \frac{1-\frac{\lambda^{m+1}}{(1+\lambda)^{m+1}}}{1-\frac{\lambda}{1+\lambda}} \mathbb{P}^{\infty}\big(\exists k\in \mathbb{N}. \widehat{\bm{\phi}}^{\bm{x}}_{\pi}(k) \in \mathcal{X}_r \mid  \bm{x}\in \mathcal{X}_0 \big)
\end{split}
\]
and consequently,
$\epsilon_1\leq v(\bm{x})\leq \mathbb{P}^{\infty}\big(\exists k\in \mathbb{N}. \widehat{\bm{\phi}}^{\bm{x}}_{\pi}(k) \in \mathcal{X}_r \mid \bm{x}\in \mathcal{X}_0\big)$. \qed
}

\textbf{The proof of Proposition \ref{pro_e2}:}

\pf{
From \eqref{ebs}, we have that
\begin{equation*}
    \begin{split}
    &v(\bm{x})\geq \frac{1}{1+\lambda}1_{\mathcal{X}_r}(\bm{x})+\frac{\lambda}{1+\lambda}\mathbb{E}^{\infty}[v(\widehat{\bm{\phi}}_{\pi}^{\bm{x}}(1))],\\
    &\frac{\lambda}{1+\lambda} \mathbb{E}^{\infty}[v(\widehat{\bm{\phi}}_{\pi}^{\bm{x}}(1))]\geq \frac{\lambda}{(1+\lambda)^2}\mathbb{E}^{\infty}[1_{\mathcal{X}_r}(\widehat{\bm{\phi}}_{\pi}^{\bm{x}}(1))]\\
    &~~~~~~~~~~~~~~~~~~~~~~~~~~~~~~~~~~~~~~~~~~~~+\frac{\lambda^2}{(1+\lambda)^2}\mathbb{E}^{\infty}[v(\widehat{\bm{\phi}}_{\pi}^{\bm{x}}(2))],\\
    &\frac{\lambda^2}{(1+\lambda)^2} \mathbb{E}^{\infty}[v(\widehat{\bm{\phi}}_{\pi}^{\bm{x}}(2))]\geq \frac{\lambda^2}{(1+\lambda)^3}\mathbb{E}^{\infty}[1_{\mathcal{X}_r}(\widehat{\bm{\phi}}_{\pi}^{\bm{x}}(2))]\\
    &~~~~~~~~~~~~~~~~~~~~~~~~~~~~~~~~~~~~~~~~~~~~+\frac{\lambda^3}{(1+\lambda)^3}\mathbb{E}^{\infty}[v(\widehat{\bm{\phi}}_{\pi}^{\bm{x}}(3))],\\
    &\ldots,\\
    &\frac{\lambda^m}{(1+\lambda)^m} \mathbb{E}^{\infty}[v(\widehat{\bm{\phi}}_{\pi}^{\bm{x}}(m))]\geq \frac{\lambda^m}{(1+\lambda)^{m+1}} \mathbb{E}^{\infty}[1_{\mathcal{X}_r}(\widehat{\bm{\phi}}_{\pi}^{\bm{x}}(m))]\\
    &~~~~~~~~~~~~~~~~~~~~~~~~~~~~~~~~~+\frac{\lambda^{m+1}}{(1+\lambda)^{m+1}}\mathbb{E}^{\infty}[v(\widehat{\bm{\phi}}_{\pi}^{\bm{x}}(m+1))],\\
    &\ldots.
    \end{split}
\end{equation*}
Thus, we can obtain that for $m\geq N$, 
\[
\begin{split}
&v(\bm{x})\geq \frac{\lambda^{m+1}}{(1+\lambda)^{m+1}}\mathbb{E}^{\infty}[v(\widehat{\bm{\phi}}_{\pi}^{\bm{x}}(m+1))]+\\
&~~~~~~~~~~~\frac{\lambda^N}{(1+\lambda)^{N+1}} \frac{1-\frac{\lambda^{m+1-N}}{(1+\lambda)^{m+1-N}}}{1-\frac{\lambda}{1+\lambda}} \times\\ &~~~~~~~~~~~\mathbb{P}^{\infty}\big(\exists k\in [0,N]. \widehat{\bm{\phi}}^{\bm{x}}_{\pi}(k) \in \mathcal{X}_r \mid  \bm{x}\in \mathcal{X}_0 \big)
\end{split}
\]
and consequently,
\[
\begin{split}
&\frac{(1+\lambda)^N}{\lambda^N}\epsilon'_2\geq \frac{(1+\lambda)^N}{\lambda^N}v(\bm{x})\\
&\geq \mathbb{P}^{\infty}\big(\exists k\in [0,N]. \widehat{\bm{\phi}}^{\bm{x}}_{\pi}(k) \in \mathcal{X}_r \mid \bm{x}\in \mathcal{X}_0\big).
\end{split}
\]\qed
}

     \textbf{The proof of Proposition \ref{pro_e4}:}
    \pf{Assume $\bm{x}\in \mathcal{X}_0$.
    According to \eqref{ebs_3}, we have that 
    \[
    \begin{split}
    &v(\bm{x})\geq 1_{\mathcal{X}_r}(\bm{x}),\\
    &\widetilde{\alpha}^{-1}v(\bm{x})+\widetilde{\beta}\geq \mathbb{E}^{\infty}[v(\widehat{\bm{\phi}}_{\pi}^{\bm{x}}(1))]\geq \mathbb{E}^{\infty}[1_{\mathcal{X}_r}(\widehat{\bm{\phi}}_{\pi}^{\bm{x}}(1))],\\  &\widetilde{\alpha}^{-2}v(\bm{x})+\widetilde{\alpha}^{-1}\widetilde{\beta}+\widetilde{\beta}\geq \mathbb{E}^{\infty}[v(\widehat{\bm{\phi}}_{\pi}^{\bm{x}}(2))]\geq \mathbb{E}^{\infty}[1_{\mathcal{X}_r}(\widehat{\bm{\phi}}_{\pi}^{\bm{x}}(2))],\\
    &\ldots,\\
    &\widetilde{\alpha}^{-N}v(\bm{x})+ \widetilde{\beta} \sum_{i=0}^{N-1} \widetilde{\alpha}^{-i} \geq  \mathbb{E}^{\infty}[v(\widehat{\bm{\phi}}_{\pi}^{\bm{x}}(N))]\geq \mathbb{E}^{\infty}[1_{\mathcal{X}_r}(\widehat{\bm{\phi}}_{\pi}^{\bm{x}}(N))].
    \end{split}
    \]
    Therefore, 
    \[
    \begin{split}
       P=\mathbb{E}^{\infty}[1_{\mathcal{X}_r}(\widehat{\bm{\phi}}_{\pi}^{\bm{x}}(N))]\leq \widetilde{\alpha}^{-N}v(\bm{x})+\widetilde{\beta}\widetilde{\alpha} \frac{(1-\widetilde{\alpha}^{-N})}{\widetilde{\alpha}-1}.
        \end{split}
    \]
   Since $v(\bm{x})\leq \epsilon'_2$ for $\bm{x}\in \mathcal{X}_0$, we can obtain 
    \[P\leq \epsilon'_2 \widetilde{\alpha}^{-N}+\frac{(1-\widetilde{\alpha}^{-N})\widetilde{\alpha}\widetilde{\beta}}{\widetilde{\alpha}-1}.\]

    If $\alpha=1$, we can obtain $P\leq \epsilon'_2 +\widetilde{\beta}N$. \qed
    }

    \textbf{The proof of Proposition \ref{pro_e3}:}
\pf{
From \eqref{ebs1}, we have that 
\begin{equation*}
    \begin{split}
    &v(\bm{x})\geq (1+\widetilde{\alpha}\widetilde{\beta}-\widetilde{\alpha})1_{\mathcal{X}_r}(\bm{x})+\widetilde{\alpha}\mathbb{E}^{\infty}[v(\widehat{\bm{\phi}}_{\pi}^{\bm{x}}(1))]-\widetilde{\alpha}\widetilde{\beta},\\
    &\widetilde{\alpha} \mathbb{E}^{\infty}[v(\widehat{\bm{\phi}}_{\pi}^{\bm{x}}(1))]\geq \widetilde{\alpha}(1+\widetilde{\alpha}\widetilde{\beta}-\widetilde{\alpha})\mathbb{E}^{\infty}[1_{\mathcal{X}_r}(\widehat{\bm{\phi}}_{\pi}^{\bm{x}}(1))]\\
    &~~~~~~~~~~~~~~~~~~~~~~~~~~~~~~~~~~~~~+\widetilde{\alpha}^2\mathbb{E}^{\infty}[v(\widehat{\bm{\phi}}_{\pi}^{\bm{x}}(2))]-\widetilde{\alpha}^2\widetilde{\beta},\\
    &\widetilde{\alpha}^2 \mathbb{E}^{\infty}[v(\widehat{\bm{\phi}}_{\pi}^{\bm{x}}(2))]\geq \widetilde{\alpha}^2(1+\widetilde{\alpha}\widetilde{\beta}-\widetilde{\alpha})\mathbb{E}^{\infty}[1_{\mathcal{X}_r}(\widehat{\bm{\phi}}_{\pi}^{\bm{x}}(2))]\\
    &~~~~~~~~~~~~~~~~~~~~~~~~~~~~~~~~~~~~~+\widetilde{\alpha}^3\mathbb{E}^{\infty}[v(\widehat{\bm{\phi}}_{\pi}^{\bm{x}}(3))]-\widetilde{\alpha}^3\widetilde{\beta},\\
    &\ldots,\\
    &\widetilde{\alpha}^m \mathbb{E}^{\infty}[v(\widehat{\bm{\phi}}_{\pi}^{\bm{x}}(m))]\geq \widetilde{\alpha}^m (1+\widetilde{\alpha}\widetilde{\beta}-\widetilde{\alpha}) \mathbb{E}^{\infty}[1_{\mathcal{X}_r}(\widehat{\bm{\phi}}_{\pi}^{\bm{x}}(m))]\\
    &~~~~~~~~~~~~~~~~~~~~~~~~~+\widetilde{\alpha}^{m+1} \mathbb{E}^{\infty}[v(\widehat{\bm{\phi}}_{\pi}^{\bm{x}}(m+1))]-\widetilde{\alpha}^{m+1}\widetilde{\beta}.
    \end{split}
\end{equation*}

Thus, we can obtain that for $m\geq N$, 
\[
\begin{split}
&v(\bm{x})\geq \widetilde{\alpha}^{m+1}\mathbb{E}^{\infty}[v(\widehat{\bm{\phi}}_{\pi}^{\bm{x}}(m+1))]-\frac{\widetilde{\alpha}\widetilde{\beta}(1-\widetilde{\alpha}^{m+1})}{1-\widetilde{\alpha}}+\\
&~~~~~~~~~~~\widetilde{\alpha}^N(1+\widetilde{\alpha}\widetilde{\beta}-\widetilde{\alpha}) \frac{1-\widetilde{\alpha}^{m+1-N}}{1-\widetilde{\alpha}} \times\\ &~~~~~~~~~~~\mathbb{P}^{\infty}\big(\exists k\in [0,N]. \widehat{\bm{\phi}}^{\bm{x}}_{\pi}(k) \in \mathcal{X}_r \mid  \bm{x}\in \mathcal{X} \big).
\end{split}
\]
Also, since $v(\bm{x})\leq \epsilon'_2, \forall \bm{x}\in \mathcal{X}_0$, we have with $m\rightarrow \infty$ that 
\[
\begin{split}
&\mathbb{P}^{\infty}\big(\exists k\in [0,N]. \widehat{\bm{\phi}}^{\bm{x}}_{\pi}(k) \in \mathcal{X}_r \mid \bm{x}\in \mathcal{X}_0\big)\\
&\leq \frac{\epsilon'_2\widetilde{\alpha}^{-N}(1-\widetilde{\alpha})+\widetilde{\alpha}\widetilde{\beta}\widetilde{\alpha}^{-N}}{1+\widetilde{\alpha}\widetilde{\beta}-\widetilde{\alpha}}. 
\end{split}
\]
\qed
}

\textbf{ The proof of Proposition \ref{upper_invariance1}:}

\pf{From  constraints $v(\bm{x})\geq \mathbb{E}^{\infty}[w(\bm{\phi}^{\bm{x}}_{\pi}(1))]-w(\bm{x}), \forall \bm{x}\in \mathcal{X}\setminus \mathcal{X}_r$ and $v(\bm{x})\geq 1, \forall \bm{x}\in \mathcal{X}_r$, we have that 
\[v(\bm{x})\geq 1_{\mathcal{X}_r}(\bm{x})+\mathbb{E}^{\infty}[w(\widetilde{\bm{\phi}}^{\bm{x}}_{\pi}(1))]-w(\bm{x}), \forall \bm{x}\in \mathcal{X},\]
where $\widetilde{\bm{\phi}}^{\bm{x}}_{\pi}(\cdot): \mathbb{N}\rightarrow \mathcal{X}$ is the trajectory to system \eqref{new_system1}. We further have that for $m\in \mathbb{N}$,
\begin{equation*}
\begin{split}
&v(\bm{x})\geq 1_{\mathcal{X}_r}(\bm{x})+\mathbb{E}^{\infty}[w(\widetilde{\bm{\phi}}^{\bm{x}}_{\pi}(1))]-w(\bm{x}),\\
&v(\bm{x})+c\geq \mathbb{E}^{\infty}[v(\widetilde{\bm{\phi}}^{\bm{x}}_{\pi}(1))]\geq \mathbb{E}^{\infty}[1_{\mathcal{X}_r}(\widetilde{\bm{\phi}}^{\bm{x}}_{\pi}(1))]\\
&~~~~~~~~~~~~~~~~~~~~~~~+\mathbb{E}^{\infty}[w(\widetilde{\bm{\phi}}^{\bm{x}}_{\pi}(2))]-w(\widetilde{\bm{\phi}}^{\bm{x}}_{\pi}(1)),\\
&\ldots,\\
&v(\bm{x})+(k-1)c\geq \mathbb{E}^{\infty}[v(\widetilde{\bm{\phi}}^{\bm{x}}_{\pi}(k-1))]\\
&\geq \mathbb{E}^{\infty}[1_{\mathcal{X}_r}(\widetilde{\bm{\phi}}^{\bm{x}}_{\pi}(k-1))]+\mathbb{E}^{\infty}[w(\widetilde{\bm{\phi}}^{\bm{x}}_{\pi}(k))]\\
&~~~~~~~~~~~~~~~~~~~~~~~~~~~~~~~~~~~~~-w(\widetilde{\bm{\phi}}^{\bm{x}}_{\pi}(k-1)),\\
&v(\bm{x})\geq \mathbb{E}^{\infty}[v(\widetilde{\bm{\phi}}^{\bm{x}}_{\pi}(k))]\geq \mathbb{E}^{\infty}[1_{\mathcal{X}_r}(\widetilde{\bm{\phi}}^{\bm{x}}_{\pi}(k))]\\
&~~~~~~~~~~~~~~~+\mathbb{E}^{\infty}[w(\widetilde{\bm{\phi}}^{\bm{x}}_{\pi}(k+1))]-w(\widetilde{\bm{\phi}}^{\bm{x}}_{\pi}(k)),\\
&\ldots,\\
&v(\bm{x})+(k-1)c\geq \mathbb{E}^{\infty}[v(\widetilde{\bm{\phi}}^{\bm{x}}_{\pi}(mk-1))]\\
&\geq \mathbb{E}^{\infty}[1_{\mathcal{X}_r}(\widetilde{\bm{\phi}}^{\bm{x}}_{\pi}(mk-1))]+\mathbb{E}^{\infty}[w(\widetilde{\bm{\phi}}^{\bm{x}}_{\pi}(mk))]\\
&~~~~~~~~~~~~~~~~~~~~~~~~~~~~~~~~~~~~~-w(\widetilde{\bm{\phi}}^{\bm{x}}_{\pi}(mk-1)).
\end{split}
\end{equation*}
Consequently, \[
\begin{split}
&v(\bm{x})+\frac{(k-1)c}{2}\geq \frac{\mathbb{E}^{\infty}[w(\widetilde{\bm{\phi}}^{\bm{x}}_{\pi}(mk))]}{mk}+\\
&~~~~~~~~~~\frac{1_{\mathcal{X}_r}(\bm{x})+\ldots+\mathbb{E}^{\infty}[1_{\mathcal{X}_r}(\widetilde{\bm{\phi}}^{\bm{x}}_{\pi}(mk-1))]}{mk}, \forall m\in \mathbb{N}.
\end{split}
\]
As $m$ approaches infinity, we conclude 
\[v(\bm{x})+\frac{(k-1)c}{2}\geq \mathbb{P}^{\infty}\big(\exists i\in \mathbb{N}. \bm{\phi}^{\bm{x}}_{\pi}(i) \in \mathcal{X}_r\mid \bm{x}\in \mathcal{X}\big).\]
Also, since $v(\bm{x})\leq \epsilon'_2$ for $\bm{x}\in \mathcal{X}_0$, we obtain 
\[\epsilon'_2+\frac{(k-1)c}{2}\leq \mathbb{P}^{\infty}\big(\exists i\in \mathbb{N}. \bm{\phi}^{\bm{x}}_{\pi}(i) \in \mathcal{X}_r\mid \bm{x}\in \mathcal{X}_0\big).\]
The proof is completed. \qed
}
\end{document}